# THE MOCHI.LABJET EXPERIMENT FOR MEASUREMENTS OF CANONICAL HELICITY INJECTION IN A LABORATORY ASTROPHYSICAL JET

SETTHIVOINE YOU, JENS VON DER LINDEN, ERIC SANDER LAVINE, EVAN GRANT CARROLL, ALEXANDER CARD,
MORGAN QUINLEY, MANUEL AZUARA-ROSALES
Aeronautics & Astronautics, University of Washington, Box 352250, Seattle, WA 98195, USA
syou@aa.washington.edu

## ABSTRACT

The Mochi device is a new pulsed power plasma experiment designed to produce long, collimated, stable, magnetized plasma jets when set up in the LabJet configuration. The LabJet configuration aims to simulate an astrophysical jet in the laboratory by mimicking an accretion disk threaded by a poloidal magnetic field with concentric planar electrodes in front of a solenoidal coil. The unique setup consists of three electrodes, each with azimuthally symmetric gas slits. Two of the electrodes are biased independently with respect to the third electrode to control the radial electric field profile across the poloidal bias magnetic field. This design approximates a shear azimuthal rotation profile in an accretion disk. The azimuthally symmetric gas slits provide a continuously symmetric mass source at the footpoint of the plasma jet, so any azimuthal rotation of the plasma jet is not hindered by a discrete number of gas holes. The initial set of diagnostics consists of current Rogowski coils, voltage probes, magnetic field probe arrays, an interferometer and ion Doppler spectroscopy, supplemented by a fast ion gauge and a retarding grid energy analyzer. The measured parameters of the first plasmas are $\sim 10^{22}$ m$^{-3}$, $\sim 0.4$ T, 5-25 eV, with velocities of $\sim$20-80 km/s. The combination of a controllable electric field profile, a flared poloidal magnetic field, and azimuthally symmetric mass sources in the experiment successfully produces short-lived ($\sim 10$ $\mu s$, $\gtrsim 5$ Alfvén times) collimated magnetic jets with $\sim$10:1 aspect-ratio and long-lived ($\sim 100$ $\mu s$, $\gtrsim 40$ Alfvén times) flow-stabilized, collimated, magnetic jets with $\sim$30:1 aspect-ratio.

*Key words:* galaxies:jets – ISM: jets and outflows – plasmas – stars: jets – stars: winds and outflows

## 1. INTRODUCTION

Understanding the interaction between plasma flows and magnetic fields remains a fundamental problem in astrophysical jets and in fundamental plasma physics. In natural cylindrical configurations, strong flows are understood to be of primary importance in astrophysical jets (De Young 1991; Meier, et al. 2001; Lovelace et al. 1993) even if it is not fully understood how the jets attain such large aspect-ratios without becoming grossly unstable (Begelman 1998; McKinney & Blandford 2009; Duran et al. 2017). Magnetohydrodynamics are thought to drive both stellar jets (Shu et al. 1994, Reipurth & Bally 2001) and relativistic jets (Lovelace & Romanova 2003), with the electromagnetic field carrying most of the energy and angular momentum in relativistic cases (Mizuno et al. 2014, Singh et al. 2016). In current-free magnetic nozzles of plasma thrusters, ion flow surfaces must somehow collimate to improve thrust efficiency (Takahashi & Ando 2017) and somehow detach from magnetic field surfaces (Fruchtman et al. 2012; Arefiev & Breizman 2005; Ahedo & Merino 2011; Olsen, et al. 2015). Axial flows have been shown to stabilize Z-pinches from current-driven instabilities (Shumlak & Hartman 1995; Shumlak et al. 2003), and also provide a mechanism for collimating current-carrying magnetic flux tubes (Bellan 2003), even though it is uncertain whether axial flows alone or with azimuthal flows can stabilize screw pinches. Recent observations show that magnetized jets travelling into transverse vacuum magnetic fields can develop axial shear flows sufficient to stabilize the jet to classical kinks (Zhang et al. 2017). Flows generated by slingshot effects (Ono, et al. 1996) during magnetic reconnection of merging compact toroids can be sufficiently sheared to prevent the cascade of magnetic perturbations responsible for toroidal field generation (Kawamori et al. 2005). Flows appear then to regulate energy partitioning during magnetic reconnection in space (Hara et al. 2011) or the laboratory (Tanabe et al. 2015). Flows appear to be crucial for energy confinement transitions in toroidal magnetic configurations even if it is still not fully understood how the appropriate flow structures appear spontaneously (Doyle et al. 2007; Diamond et al. 2005; Spence et al. 2009). In natural spherical configurations, sustained non-axisymmetric flows appears to be necessary for resolving the dynamo problem and the origin of cosmic magnetic fields even if field generation remains to be demonstrated experimentally at the relevant regimes (Kulsrud & Zweibel 2008; Widrow 2002; Cooper et al 2014).

In all these cases, there appears to be a spontaneous emergence of flows or magnetic structures, suggesting a form of self-organization in plasmas that requires careful consideration of the coupling between magnetic fields and flows. A fundamental conjecture in self-organization is the invariance of a global property during a relaxation process: for example, a system reduces energy (Taylor 1974; Moffatt et al. 2013), or equivalently, maximizes entropy (Steinhauer & Ishida, 1997), but conserves the total helicity of a vector field. Above a critical threshold, the system then forms a large-scale structure of that specific vector field. The critical threshold is the eigenvalue of the flux conserving volume (Taylor 1974, Bellan 2000). Hydrodynamic flow helicity models (Moffat 1969; Moffatt et al. 2013) have been applied to the forecasting of tornadoes (Rasmussen & Blanchard 1998), knotting of DNA (Ernst & Sumners 1990), entanglement of polymers (Lehn 1995), and wing-tip vortices (Elimelech et al. 2013). Magnetic field helicity models (Brown et al. 1999) are the foundation of theories for the origin of cosmic magnetic fields (Kulsrud & Zweibel 2008), astrophysical jets (Li et al. 2006), solar coronal loops (Mahajan et al. 2001), and toroidal magnetic confinement concepts (Brown et al. 1999; Bellan 2000; Steinhauer 2011). More recent theories bridge the gap between self-organization of hydrodynamic flows and magnetostatic configurations by supposing the more fundamental topological invariant is the helicity of canonical momentum (for short, we call this type "canonical helicity", if also known as generalized vorticity (Turner 1986), self-helicity (Steinhauer & Ishida 1997), generalized helicity (Oliveira & Tajima 1995), or fluid helicity (Avinash 1992)). The strength of this approach is the ability to predict the structures of two-fluid plasmas, for example, in isolated toroidal configurations with significant flows, angular momentum and pressure gradients (Steinhauer & Ishida 1998) or isolated double-Beltrami states in complex multi-fluids with dust (Mahajan & Lingam 2015). Another strength is the ability to predict the threshold for bifurcation in compact torus formation from the merging of counter-helicity spheromaks (You 2014; You 2012).

In the most general form, canonical helicity evolution is also valid in kinetic regimes and relativistic systems (You 2016). This mathematical result opens the door to considering self-organization in realistic regimes beyond the simplest neutral (Moffat 1969), magnetohydrodynamic (MHD) (Taylor 1974) or at best, barotropic multi-fluids (Mahajan & Lingam 2015; Steinhauer et al. 2001). The theory (You 2012; You 2016) predicts conversion of magnetic helicity to flow helicity and vice versa; changes in helicity that are lower than energy changes in weak density gradients but are larger than energy changes in steep density gradients; a one-to-one relationship between the plasma species and canonical flux tubes; and natural collimation of canonical flux tubes with significant flow vorticities (Lavine & You 2017). These results suggest that, in a cylindrical configuration, appropriate enthalpy boundary conditions could establish sufficient axial and azimuthal shear flows to stabilize a collimated plasma jet over long aspect-ratios. Therefore to verify this prediction, and with the conjecture that this is applicable to astrophysical jets (Lavine & You 2017; Ryutov et al. 2001; Lebedev et al. 2002; Bellan et al. 2005), the LabJet configuration of the Mochi project is designed with a reasonable (and experimentally feasible) model of an astrophysical jet launching system. The experimental facility is named Mochi for the Japanese rice cake. There is one possible "back-cronym": Measurements Of Canonical Helicity Injection.

This paper introduces the concept of the Mochi.LabJet experiment (Section 2), the design (Section 3), the hardware (Sections 4, 5), initial diagnostics (Section 6) and measured initial plasma parameters (Section 7). The nomenclature of the hardware follows an object-oriented convention with a period separating the objects, such as Mochi.Labjet or LabJet.CoreGun to describe the hierarchy and relationship of hardware items. Section 8 then summarizes and concludes.

## 2. CONCEPT

There are generally two classes of laboratory simulations of astrophysical jets: high energy density hydrodynamic jets driven by lasers or magnetic pinches (Remington et al. 2006; Bouquet et al. 2010; Remington et al. 2006; Lebedev et al. 2002) and medium energy density magneto-hydrodynamic jets driven by plasma guns (Z pinches (Shumlak et al. 2003), screw pinches (Hsu & Bellan 2002, 2005; Bellan et al. 2005)).

High energy density experiments exploit recent advances in pulsed power technology and diagnostic techniques developed for inertial fusion research in attempts to replicate or approach actual astrophysical dimensionless numbers (Bouquet et al. 2010) but without replicating actual accretion disk boundary conditions. The primary motivation is to study the detailed effects of high Mach number, radiative cooling and shocks on the collimation and lengthening of jets by careful control of experimental parameters and sophisticated diagnostics. Laser driven experiments such as the Omega facility (Boehly et al. 1997; Foster et al. 2005), NIF (Hogan et al. 2001), and LMJ in France (Besnard 2007) typically use several kJ lasers to ablate a pusher material that creates a strong shock wave into a piece of plastic foam. The high >100 Mbar pressures obliterate the foam into a plasma which squeezes through a micron size conical aperture designed to collimate the jet as it expands into a large vacuum chamber. Minuscule ~1 mm jets shoot out at ~100 km/s with a lifetime of 1-10 ns at densities of ~$10^{25}$ m$^{-3}$, typically made of ~~high Z~~ particles like CH$_2$ with mass densities of $10^{-4}$ g/cm$^3$ at temperatures of a few eV. Laser driven experiments are purely hydrodynamic or are radiation-dominated and do not yet have internal magnetic fields, although some have been collimated with an external magnetic lens arrangement (Chen et al. 2014). Alternative experiments on the MAGPIE facility (Lebedev et al. 2002, Ampleford et al. 2008) ablate metal wires suitably arranged in a conical fashion. The mega-Ampère currents turn the wires into current-carrying Z-pinch plasmas that attract each other and merge on the axis of symmetry. The conical or purely radial arrangement introduces an axial pressure gradient that drives a jet along the axis. Jets ~few cm long with ~5:1 aspect ratios propagate at velocities of ~200 km/s, estimated $M_{jet}$ ~ 30 with $10^{25}$ m$^{-3}$ densities into vacuum (very high density ratio). Radiation cooling appears to play an important role in the collimation and extension of the jet. Dimensionless numbers begin to approach astrophysical jet conditions, except for the extreme aspect ratios. It is currently unclear if longer aspect ratios can be achieved before instabilities destroy the jet, without magnetic fields and shear flows (although some azimuthal flows have been introduced (Ampleford et al. 2008)).

Medium energy density experiments attempt to model some of the global behavior of astrophysical jets by setting appropriate boundary conditions and sufficiently small or large dimensionless numbers but without trying to achieve actual astrophysical values. Sufficiently small or large means for example that a laboratory Lundquist number $S \sim 10^3 - 10^5$ is not different on a global scale to $10^{10}$, because there is not much difference if the resistive term in MHD is $10^5$ times smaller or $10^{10}$ smaller than the Lorentz terms.

The ZaP experiment (Shumlak et al. 2003) has produced long-lived, ~30:1 aspect-ratio Z-pinch plasmas with axial shear flows strong enough to stabilize gross instabilities (kink, sausage, Rayleigh-Taylor). The cylindrical coaxial plasma gun consists of a short formation-acceleration region which becomes a long ~1 m inertial propagation region. A pre-formed annular plasma slug exits the acceleration region with the inner plasma remaining attached to the few cm diameter center cathode, while the outer plasma travels along the outer anode all the way to the other end at ~ 1 m until the plasma assembles into a long Z-pinch on the axis of symmetry. Inertia maintains the axial shear flow in the plasma column connecting the cathode to the outer end anode. Banks of 28-46 kJ drive ~200 kA of current with a plasma quarter cycle duration of ~30 $\mu s$ giving ~2 T azimuthal pinch magnetic fields. Velocities of ~50 km/s are achieved with electron densities measured at ~$10^{22-23}$ m$^{-3}$ inside the hydrogen plasma column and temperatures estimated from force balance at ~200 eV. The experiment demonstrates that long aspect-ratios can be achieved with sufficient shear flow stabilization on Z-pinches but does not attempt to simulate astrophysical jet boundary conditions with a magnetized, rotating accretion disk.

The Caltech astrophysical jet experiment (Hsu & Bellan 2002, Hsu & Bellan 2005, Bellan et al. 2005) is based on coaxial gun technology with similar experimental parameters to the ZaP experiment but different boundary conditions. The Caltech apparatus has observed collimation effects from strong $\vec{J} \times \vec{B}$ forces in an experimental setup which more closely reproduces the astrophysical jet-accretion disk system with a planar plasma gun (Hsu & Bellan 2002, Hsu & Bellan 2005, Bellan et al. 2005; You et al. 2005). The single line-tied screw pinch column achieved aspect-ratios ~10:1 before large Kruskal-Shafranov instabilities twist the column and break off to form a new (spheromak-like) configuration (Hsu & Bellan 2003). Although large velocities ~50-70 km/s were measured, the setup presumably did not establish shear flows inside the column sufficiently strong to stabilize the current-driven instabilities and the 30 cm long jet lasted ~5-10 $\mu s$ before detachment and disappearance.

Because the circulation of $\vec{J} \times \vec{B}$ forces is ultimately responsible for collimating the jet and for driving strong axial flows (Bellan 2003), the next simplest step in simulating an astrophysical jet-accretion disk system in the laboratory is to tailor the current profile $J_z(r)$ in a screw-pinch jet. Therefore our LabJet configuration of the Mochi device uses three concentric annular electrodes threaded by a poloidal magnetic field (Figure 1). Rotating electrodes would be impractical in the laboratory because millions of revolutions per minute would be necessary to produce the large electrical currents, and assuming $u_\theta = -E_r/B_z$, the electrodes establish a radial electric field roughly equivalent to an accretion disk rotating azimuthally in a poloidal magnetic field. The first motivation for using multiple concentric electrodes is to approximate accretion disk rotation velocity profiles by tailoring radial electric field profiles, The second motivation is to separately control the enthalpy $h = 1/2 \, nmu^2 + nq\phi$ boundary conditions ($n$ is the number density, $m$ the species mass, $q$ the species charge, $u$ the species fluid flow velocity, $\phi$ the electrostatic potential) on the skin region of the jet (radii near the edge of the tube) and the core region of the jet (smaller radii closer to the axis of the tube). This effectively provides an experiment control knob on the current profile $J_z(r)$ in the column: varying the electrode potentials will change the enthalpy (electrostatic potential plus kinetic energy) and therefore the canonical helicity injection rate at the edge and the core of the column

$$\frac{dK_\sigma}{dt} = 2 \int \boldsymbol{R} \cdot \boldsymbol{\Omega} \, dV - 2 \int h \, \boldsymbol{\Omega} \cdot d\boldsymbol{S} - \int \boldsymbol{P} \times \frac{\partial \boldsymbol{P}}{\partial t} \cdot d\boldsymbol{S} \quad (1)$$

where $K_\sigma = \int \boldsymbol{P} \cdot \boldsymbol{\Omega} \, dV$ is the relative canonical helicity (ignoring reference field subscripts and boundary motions) of the canonical momentum $\boldsymbol{P} = nm\boldsymbol{u} + nq\boldsymbol{A}$ of the species $\sigma$, $\boldsymbol{\Omega} = \nabla \times \boldsymbol{P}$ is the canonical vorticity and $\boldsymbol{R}$ represents dissipative forces (You 2016). Changing the edge vs core canonical helicity injection rate provides azimuthal shear rotation $u_\theta(r)$ and introduces sheared mass injection $u_z(r)$ into the jet during and after formation. The design therefore provides control of the radial current profile $J_z(r)$ and, indirectly, velocity profiles $u_z(r)$ and $u_\theta(r)$ in a setup which simulates the astrophysical jet-accretion disk system. We also pay careful attention to establish an azimuthally symmetric mass source at the boundaries to minimize the effect of asymmetries on jet launching and late-stage fuelling (discrete gas holes as in the Caltech experiment anchor the plasma to the holes and prevent free rotation). A fast poloidal and toroidal shear flow can then be established, and if sufficiently strong, should stabilize the long jet to current-driven instabilities.

If jets obey magnetohydrodynamic (MHD) scaling (Ryutov, et al., 2001), then times should scale as $a\sqrt{b/c}$, velocities as $\sqrt{b/c}$, and magnetic fields as $\sqrt{c}$, where $a$ is the ratio of astrophysical jet radius to laboratory jet radius, $b$ is the ratio of mass densities, and $c$ is the ratio of plasma pressure. To within a factor of 3-10, the parameters of the Mochi.LabJet experiment (Table 1) match protostellar jet parameters (Hartigan, et al., 2007; Hartigan & Morse, 2007) (radii of $10^{15}$ m, densities $10^{17}$ kg/m$^3$, pressures $10^{-9}$ Pa, time scales $10^4$ yrs, velocities 300 km/s, and magnetic fields $10^{-7}$ T). However, in nature, astrophysical jets vary widely in physical parameters *and* vary widely in dimensionless numbers (Ryutov et al. 2001; Lebedev et al. 2002; Bellan et al. 2005). Yet jets appear to share features such as central objects, rotating accretion disks, and helical magnetic fields. Boundary conditions may therefore play a more important role than exact dimensionless numbers in the launching, collimation, and stabilization mechanisms. The spontaneous appearance of a large macroscopic feature from given boundary and initial conditions, irrespective of the detailed mechanisms leading to the final state, is a feature of self-organization. Given that a laboratory experiment cannot reproduce all the dimensionless numbers simultaneously, the Caltech experiment and the Mochi experiment described here use the simplest feasible design of an accretion disk around a central engine with magnetic fields. The Mochi experiment builds on the Caltech experiment by adding another electrode to tailor the radial electric field profile and imposing azimuthally symmetric gas slits. The machine parameters are presented in Table 1a with details presented in Section 3-5. The plasma parameters are presented in Table 1b. They were determined from the first measurements (Section 7) of the initial set of diagnostics (Section 6).

## 3. DESIGN

The Mochi.LabJet plasma gun was built in-house and consists of 4 sub-assemblies (Figs. 2, 3): the inner electrode assembly, the middle electrode assembly, the outer electrode assembly, and the high-voltage connection assembly. The combination of inner electrode and outer electrode constitutes a single "core" plasma gun (CoreGun) designed to drive the core of the current-carrying magnetic flux tube, and the combination of the middle electrode and outer electrode constitutes an independent "skin" plasma gun (SkinGun) designed to drive the skin layer of the current-carrying magnetic flux tube. The intention is to provide independent control of the enthalpy profile imposed at the foot of a jet (here expressed as an electrostatic potential profile) and therefore the canonical helicity injection profile across the cross-section of the jet.

The inner electrode assembly (Fig. 2b) is made of two electrically-connected concentric tube sub-assemblies (1" and ½" outer diameter) that terminate into two annular OFHC (Cu 101 H02) copper electrodes (6cm diameter disc and 7.25cm diameter cup). Neutral gas injected through four ¼" compression tube unions at one end flow along the annular volume bounded by the two concentric tubes. Because the axial distance is much longer than the radial and azimuthal dimensions, the gas will emerge azimuthally symmetric from the 2mm slit between the copper disc and cup, at a radius of 3.05cm. The inner electrode assembly is attached to the middle electrode assembly via an o-ring quick disconnect feedthrough mounted on a high-voltage 2.75" CF ceramic break, aligned with a machined ceramic insulating insert.

The middle electrode assembly (Fig. 2c) consists of a 7" re-entrant stainless steel capped tube and a concentric 2.5" stainless steel tube welded to a 10" CF port terminating on two annular OFHC copper electrodes (7" diameter annulus and 9.6cm diameter ring). Gas is injected directly to the back of the annulus copper electrode through twelve ¼" stainless steel tubes. The gas makes radial and azimuthal transits from the discrete holes to fill up the plenum between the annulus, ring and re-entrant tube end-cap, and emerges azimuthally symmetric from the 2mm slit at a radius of 4.9cm. A custom alumina insert is torr-sealed to the 7" diameter copper annulus to minimize arcing between the middle and outer electrodes. The two assemblies are mounted to a 10" CF flange on the vacuum chamber end-dome door via a high-voltage ceramic break. The re-entrant adapter houses the bias magnetic field solenoid, made of 4mm square magnet wire wound $11 \times 9$ times around a 6"diameter phenolic support (1 mH, 1.5 Ω).

The outer electrode (Fig. 2d) is an annulus of 90cm outer diameter and 10.4cm inner diameter, made of three 4.83mm thick OFHC copper annular sheets mated to two stainless steel annular gas plenums, giving an "outer inner" and an "outer outer" 20mm gas slit at a radius of 19.1cm and 40.1cm, respectively. These radial gas slit locations were chosen to face the "inner" and "middle" gas slits on the inner and middle electrodes to promote gas breakdown when following the vacuum bias poloidal magnetic field lines. Each plenum is built from stacked, torr-sealed, stainless steel rings and pierced annular sheets mated to sixteen ¼" stainless steel gas delivery tubes. The design forces the gas to zig-zag long distances azimuthally and radially before emerging through the slits in between the copper annuli for an azimuthally symmetric gas source. Eight circular cutouts provide accessible lines-of-sight for the 3.375" viewports on the end-dome door of the chamber. The outer electrode assembly is attached to stainless steel rods welded to the hinged end-dome door with ¼" gas tubes passed through o-ring feedthroughs connected with compression unions to the plenums.

The complete LabJet assembly is mounted on a hinged end-dome door (Fig. 2a, 3) with double o-ring grooves (for differential pumping if needed) of a custom, electro-polished, stainless steel, 60" (1.5 m) diameter spherical vacuum chamber. The chamber, manufactured to specification by MDC Vacuum, has numerous ports: two 14" CF flanges at the north and south poles to house a Sumitomo Marathon CP-12 cryopump (3600 L/s giving a typical chamber base pressure of $\sim 3 \times 10^{-7}$ torr without baking nor surface conditioning) and extra hardware; six 8" CF glass viewports along the equator, aligned to the origin of the spherical chamber; thirty 2.75" CF glass viewports along the tropics and in front of each end-dome that point radially in a vertical plane (azimuthal plane) for diagnostics; sixty four 3.375" CF glass viewports regularly distributed around, and pointing to the origin of, the sphere dedicated to tomographic spectroscopy (all ports are laser-aligned to within $\pm 1^o$ and $\pm 0.030$" tolerance); one 41" wire-seal port to mount the Mochi.SpiderLeg plasma gun (renamed Driven Relaxation eXperiment DRX gun provided by Los Alamos National Laboratory (Hsu & Tang 2008). This gun is made from two planar coaxial electrodes with 8 pairs of gas holes driven by a custom capacitor bank power supply (sixty NWL 12094 capacitors, 240 $\mu$F, 10 kV, 12 kJ total) in front of a double-wound solenoid driven by a custom capacitor bank (48 mF, 450 V, 4.5 kJ) to create a bias poloidal magnetic field. The Mochi.SpiderLeg gun can serve as comparison for the Mochi.LabJet gun to highlight the influence of three electrodes and azimuthally symmetric gas slits over two electrodes and eight pairs of discrete gas holes (Fig. 3). It is possible to operate both guns simultaneously for studies of jet merging or jet impact on targets to simulate astrophysical jet impacts on vacuum magnetic fields, on neutral gas or on plasma clouds. The configuration can also be adapted to flow investigations in toroidal geometry by forming and merging spheromaks from planar plasma guns to investigate compact torus formation, compression, and magnetic reconnection.

## 4. POWER SUPPLIES

The LabJet.Core and LabJet.Skin plasma guns are powered by identical pulsed power supply units (PSU, Fig. 4). A PSU contains five capacitors (GA model 33838, 120 $\mu$F, 10 kV, 6 kJ each) connected in parallel with 1" wide silver-plated copper bus bars and connected to a custom-design tower that houses the ignitron (EEV BK178, size D, 20 kV, 100 kA, 200 A.s/pulse). The bus bars can be readily replaced to vary inductances for a pulse-forming network. The co-axial custom tower (Fig. 4d) was developed as a cost-effective and reliable connection that minimizes inductive impedance. The forward current assembly connects the ignitron to the capacitor bank and the high-voltage (HV) coaxial cables. The return current assembly consists of eight brass sheets acting as a co-axial return path. Despite three times the resistivity, brass was selected to reduce cost because the resistive impedance is only increased by a square root of three over copper when considering the high-frequency (10 kHz-1 Mhz) limit of skin effects appropriate for our experiments. The ignitrons are triggered through a transformer by a custom fast drive circuit (Chaplin & Bellan, 2013). Each PSU is connected to the LabJet gun from the top of the ignitron tower with eight parallel RG-217U coaxial HV cables (15 ft, 50 Ω impedance each) with high-power RC snubbers (7.2 Ω, 0.05 $\mu$F) for impedance matching. Each capacitor bank is charged with a Spellman SL10N300 HV power supply capable of delivering 30 mA of current at negative 10 kV, and remotely operated via custom optical isolation circuits (Figs. 5-6, Section 5). The output from the charging supply is switched by a "charge" Ross high-voltage relay itself switched by a custom optical relay circuit (Figs. 5-6, Section 5). An additional optically-controlled "dump" Ross relay (modified to be normally closed) connects the high-voltage terminal of the capacitors to four 400 Ω silicon carbide high-power resistors combined into 200 Ω to safely dump the stored energy over an RC discharge time of 120 ms. Both power supplies are housed in wheel-mounted enclosures for mobility. All panels of the enclosure are interlocked with a latching relay circuit. If any panel is removed while the power supply is powered, the dump relay is immediately closed to de-energize the system. The latching relay circuit must be reset to continue normal operation, acting as a fault acknowledgement system. Manual switches, displays and optical fiber connectors are mounted on a 3D printed control panel on the front of the PSU enclosure.

## 5. CONTROL AND DATA ACQUISITION

The Mochi experiment is operated remotely via full optical isolation for safety and convenience (Fig. 5). The optical signals are of three types: long-duration (DC) signals that control relay actuation, frequency-modulated pulses (0-400 Hz) that adjust bank charging voltages, and fast ($< \mu s$) single pulses that trigger the equipment and diagnostics in the desired sequence. Feedback signals (for measuring charging voltage and current) are also optically transmitted back to digitizers by pulse-width modulation. The control and data acquisition equipment consists of off-the-shelf National Instruments timing and digitizer cards with custom opto-electronic transmitters and receivers, manipulated by a custom LabView PC graphical user interface (GUI). The control system is based on two PXI-6602 timing cards with 16 pulse generators clocked at 80 MHz for fast triggering or adjustment of the experimental equipment and diagnostics, and 24 digital DC outputs from three PXI-6133 multifunction IO cards for controlling relay actuation. The data acquisition system currently has 128 digitizing channels, of which 96 channels are dedicated to magnetic (Bdot) probe arrays taking 40000 samples at 50 Ms/s with 12 bit resolution and 26 MHz bandwidth, made up of three NI-5752 analog-digital adapters mated to PXIe-7962R field-programmable array (FPGA) cards programmed in-house for control of the data streams. The FPGA read out the samples as 16 bit words for a theoretical data throughput of 3.2GB/s, but the limited onboard DRAM required a custom FPGA code (von der Linden 2017) resulting in a throughput of 3GB/s and a total acquisition time of 800 µs after which the overflow data fills the onboard block memory. Five more cards (160 channels) are awaiting implementation. The next 24 digitizing channels take 2M samples at 2.5 Ms/s with 14 bit resolution at 1.3 MHz bandwidth, made up of three PXI-6133 cards housed in a PXI-1042Q chassis slaved to the PXIe-1082 chassis housing the FPGA boards. A pair of PXIe-6672 and PXI-6653 synchronization cards take care of overall synchronisation and communication to the PC. The remaining 8 channels are dedicated to HV probes for measuring gun voltages and outputs of a HeNe interferometer, made of two Tektronix 2024B digital storage oscilloscopes that take 2500 samples at 2 Gs/s with 200 Mhz bandwidth and remotely operated by a PC.

The custom opto-electronic converters consist of transmitters, fast receivers, DC receivers, optical tachometers and reverse optical tachometers (Fig. 6). The transmitters (Fig. 6a) efficiently convert either fast single electrical pulses, long-duration DC voltages or square-wave voltages into the equivalent optical signals. The input voltage biases the base of the single transistor, closing the switch and allowing the collector-emitter current to power the HFBR-1412 optical transmitter. Our circuit has an average delay of 310 ns, a jitter of 2.8 ns and only requires 0.1 W in the on-state and negligible power in the off-state (c.f. 1.05 W in the on-state and 0.33 W in the off-state for the recommended drive circuit for the HFBR-1412).

The fast receivers (Fig. 6b) converts a 100 µs light pulse into an electrical TTL pulse for triggering the capacitor bank switches and diagnostics in the chosen sequence. A light pulse causes the voltage on the output of the HFBR-2412 receiver Q1 to drop to 0.5 V, which turns off the common-emitter amplifier transistor Q2 and therefore inverting the voltage drop into a voltage rise at the collector of Q2. The emitter-follower transistor Q3 provides current gain for the TTL out. A modified Baker's clamp (Chaplin & Bellan 2013) reduces the switching time of this fast receiver circuit by reducing the time to turn off Q2 to 250 ns. The clamp allows current to flow from the base into the collector-emitter path, thereby reducing the buildup of charge on the transistor base and preventing saturation.

The DC receiver circuit (Fig. 6c) controls long time-scale switching such as actuating relays for AC circuits that power high-voltage relays in the capacitor bank charge or dump circuits. The DC receiver circuit has the same common-emitter amplifier transistor Q2 and emitter-follower transistor Q3 as the fast receiver circuit, but omits the modified Baker's clamp because high speed switching is not required. As the receiver Q1 receives light, the emitter-follower provides current gain for charging the MOSFET Q3 gate. Once the gate charge has reached turn-on threshold, the source drain current of Q3 can drive higher power components such as a reed DC relay or SCR.

The optical tachometer converts optical pulses of a given frequency into a steady, proportional voltage. The circuit consists of the fast receiver circuit coupled to a frequency-to-voltage converter (Fig. 6d) adapted from the recommended circuit for the TI LM2907 IC chip. A comparator triggers a charge pump internal to U1 each time the input signal crosses a set voltage level. The charge pump charges capacitor C1 by drawing current across resistor R4. The filter capacitor C2 converts the pulsed-voltage drop across R4 to a steady-state voltage, proportional to the input frequency. A transistor in the LM2907 amplifies the output current to buffer the load impedance from the receiver circuit. Each optical tachometer is calibrated for an input frequency range between 0-400 Hz and output voltages 0-10 V (Fig. 6e) showing a linear relationship better than 0.99. The tachometer provides the demand voltage (0-10V) to the charging power supplies needed to remotely set the charge voltage (0-10 kV) of the experiment's capacitor banks.

The reverse optical tachometer converts the capacitor bank voltage measured across a voltage divider into optical pulses, with a pulse-width proportional to the voltage. The circuit (Fig. 6f) consists of an instrumentation amplifier, a voltage-to-frequency converter circuit, and a transmitter circuit. The instrumentation amplifier INA118 inverts and measures the negative capacitor bank voltage through a 1:1000 voltage divider. The voltage-to-frequency converter LM331 (U2) repeatedly shorts $f_{out}$ to ground with a frequency proportional to the input voltage. The charge and discharge cycle of the the reference capacitor C3 sets the frequency: each time the voltage of C3 drops below the input voltage, an internal comparator triggers a one-shot timer, biasing a transistor, shorting the fast receiver input, and connecting a current source to C1. Once the one-shot timer turns off, C3 discharges until the capacitor voltage drops below the input voltage. Each reverse optical tachometer is calibrated for an input voltage range of 0-10 V with output pulse-widths ranging from 10 µs to seconds (Fig. 6g). The pulse width increases exponentially with increasing voltage for voltages below -2V and faster than exponential for voltages above -2 V. To handle this variation in pulse-width, the custom LabView control software interpolates calibration measurements with cubic splines and measures pulse widths no longer than 100 ms to take multiple measurements per second. The same custom opto-electronic fast receivers then convert the optical pulses to voltage pulses for the PXI cards.

A flexible experiment control GUI (von der Linden et al. 2017) based on the producer-consumer design pattern of LabView allows operators to manipulate all the experiment settings in real time, load previous shot settings, and store the acquired data in a hierarchical MDSplus tree (Stillerman et al. 1997). The experiment control application accesses the storage through the MDSplus LabView interface (Manduchi et al. 2013) which leverages the object-oriented and multi-threaded features of MDSplus.

## 6. DIAGNOSTICS

Voltages are measured with two Tektronix P6015 1000× high voltage probes connected to a four channel Tektronix TDS2024B digitizing oscilloscope. One probe measures the outer skin gun voltage and one measures the inner core gun voltage. Custom opto-isolated high voltage probes have been built to replace the Tektronix probes but await comprehensive tests. Currents are measured with passively-integrated Rogowski coils calibrated to a commercial Pearson current monitor. Droop correction is numerically applied with integration of the digitized signals. One Rogowski coil is located at the inner high voltage ceramic break of the inner core gun to measure the inner core gun current. A second Rogowski coil is located around the outer high voltage ceramic break to measure the total current going through both core and skin guns. Substracting the two signals gives the current flowing through the middle skin gun.

There are three Bdot magnetic probe arrays (Fig. 7). Each linear array consists of 135 inductor chips (Coilcraft 1008CS-472XGLB, 2mm×2mm×3mm) grouped into 45 clusters of 3D x, y, z orientations inside a cylindrical, 1m long stalk made of an 8mm OD alumina outer sheath over a 6mm OD stainless steel electrostatic tube shield. The 45 clusters are grouped into three sections: a middle, high spatial resolution section with 25 clusters separated by 11mm; near and far sections at lower spatial resolution with 10 clusters separated by 33mm. The grouping is such that the high resolution middle section intercepts the plasma jet. The three arrays are currently inserted into the vacuum chamber at different axial distances in front of the LabJet gun. Other positions are possible. For example, the three arrays could be inserted all in the same plane parallel to the plasma gun electrodes but at different angles to measure azimuthally asymmetric magnetic fields, rotating kinks, displaced jets, or spheromaks formed in the experiment. The 12bit, 50Ms/s, 26MHz bandwidth, 96 channel data acquisition system acquires the Bdot probe signals (Section 5). The operator chooses how the 96 channels are individually connected to the 405 inductor chips. This also leaves room for future increases in digitizing channels. The signals are carried to the digitizers by double-shielded, twisted-pair CAT-7 cables. The CAT-7 cables combine the inductive pickup rejection of twisted pair cables with capacitive pickup shielding. A custom, Pi-type, load-balanced, switchable 96 channel attenuator printed circuit board attenuates the signal amplitudes by 0-76 decibels in increments of 6, 10, and three 20 dB. The circuit is load balanced so that capacitive and inductive pickup rejection is further enhanced by the common mode rejection of differential digitization. The magnetic probes are calibrated for frequency response, orientation errors, alignment errors and pickup.

An unequal path length, heterodyne, quadrature HeNe interferometer (Kumar & Bellan 2006; Card 2017) provides line-integrated electron density measurements (Fig. 8). The 632.8nm, 10mW source laser has been measured to have two resonant modes with a visibility curve periodicity of ~34 cm allowing the reference beam to remain on the optical table while the scene beam of the setup travels several metres (about 12 lengths of the visibility periods) to and from the vacuum chamber for a double pass through the plasma. The hybrid Michelson-Mach-Zehnder setup uses an 80 MHz acousto-optic modulator (Isomet 1205C-2 with 532B-2 driver) as primary beamsplitter. The signal from a silicon photodiode with custom amplifier is demodulated through a custom RF mixer-splitter against this frequency to obtain the quadrature components of the phase angle difference. The two components are digitized with a Tektronix TDS2024B digitizing oscilloscope at 2 GS/s over the discharge duration (~100 $\mu$s).

Optical line spectra are measured with a spectroscopic setup based on a 1m focal length, f/8 monochromator (Horiba Yvon 1000M) with a 2400 lines/mm diffraction grating recorded on a PI-MAX3 ICCD camera. A 94 channel, 10 m long fiber-bundle has been constructed in house from 100 $\mu$m core multi-mode fibers, arranged in three vertical lines of 32 fibers at the matching-optics side, breaking out into twelve separate bundles of 8 fibers arranged in a line on the collimating-optics side. The complete bundle is very cost effective at about \$27 per channel, mainly due to savings in the bundle breakout and optical mounts, designed and printed in house with an off-the-shelf 3D printer, collimating lenses sourced from laser pointer parts, and protection tubing sourced from local gardening stores. The twelve bundles are mounted onto twelve 3.375" BK7 viewports of the spherical vacuum chamber, selected to observe the plasma from viewpoints regularly distributed in a toroidal and poloidal plane around the center of the sphere (Fig. 9). The fiber-bundle design and construction is also a prototype for building another with an order magnitude larger number of viewchords to take up all 64 3.375" viewports. The arrangement of these viewports is optimized for tomographic inversion of line-integrated spectra for reconstruction of 3D plasma flows and 3D plasma intensities (You et al. 2010; Balandin et al. 2012).

High-speed imaging (of the spectral lines or of the plasma directly when not mounted on the monochromator) is carried out with a Princeton Instruments PI-MAX3 1024i ICCD camera capable of taking two sequential frames at 1024×1024 pixels with 16 bit depth, typically with 1 $\mu$s exposure and 3-10 $\mu$s inter-frame time. The interframe time is limited by the phosphor intensifier decay time with a P46 phosphor decay time to 10% of 2 $\mu$s. A fast-frame "movie" CMOS camera (Specialized Imaging Kirana model) capable of taking 180 sequential 924x768 10bit frames at 5 million frames per second can be borrowed courtesy of the University of Washington Student Technology Fund.

A fast ion gauge measures the neutral particle density with 2 $\mu$s time-resolution. This gauge is used to fine tune the symmetry in neutral gas injection prior to breakdown: the various gas valves can be triggered at different times to have an azimuthally-symmetric neutral gas cloud just in front of the electrodes. During the plasma shot, the gauge is moved far away from the plasma and turned off to protect the electronics. Modifications are planned to protect the gauge during a plasma shot to enable tracking of the evolution of neutral densities in the chamber during a discharge. A retarding-field, gridded, energy analyzer has been built to measure the particle energy distribution (Quinley 2015). Courtesy of Pohang Institute of Technology (CDET-800M16WY-01), an 8-channel radio-frequency (RF) spectrometer measures RF waves from the plasma in the range 40-180 MHz to characterize ion dynamics during magnetic reconnection.

## 7. FIRST PLASMAS

The experiment achieved first plasma in Nov. 2015. The commissioning phase of the hardware was nominally completed in Aug. 2016 with a first operational campaign lasting until Nov. 2016 to determine the first plasma parameters while implementing the primary diagnostics (current, voltage, interferometer). The commissioning included testing of all major power supplies, control hardware, and data acquisition. Several crucial issues were resolved. Commercial pulse valves did not supply sufficient gas for reliable breakdown, so ten custom fast gas valves were built to deliver dense gas close to the electrodes (leaving the rest of the vacuum chamber at base pressure). A ten-channel pulsed power supply was also constructed to drive the fast gas valves. The FPGA-based digitizers exhibited unexpected resonances at ~50 kHz discovered during frequency calibrations, due to LC coupling between ADC chips and National Instruments FPGA inputs, so numerical filtering was applied. Arcing between the middle and outer electrodes during the discharge often destroyed the symmetry in jet formation or syphoned current away from the plasma jet when symmetry was maintained, so a long teflon insulating sheet was rolled around the middle electrode re-entrant port extending out the gap between the middle and outer electrodes.

After fully charging all the banks, the sequence of a typical shot (Fig. 10) begins with triggering the bias magnetic field at typically $t = -9$ ms ($t = 0$ corresponds to the trigger time of the ignitrons in the main pulsed power supplies). The exact time can be chosen to correspond to the peak current of the measured RLC pulse in the solenoid or shifted to the desired value of the current before or after the peak time to decrease the bias magnetic field strength. The RLC pulse is a critically-damped pulse with a time scale in the 10 ms range and can be considered constant on the ~100 $\mu s$ time scale of the plasma discharge (Fig. 10a). Next, gas is puffed into the vacuum chamber from the ten fast gas valves at typically $t \sim -3$ ms to -4 ms for hydrogen or $t \sim -13$ ms to $-18$ ms for argon (Fig. 10b). The time is estimated with the speed of sound of the neutral gas and fine-tuned for each gas valve with $\mu s$ time-resolved gas pressure measurements of the fast ion gauge. Fine-tuning improves the symmetry of gas injection and times for peak gas density directly in front of the electrodes without the need to pre-fill the chamber. The base chamber pressure is typically ~3x10$^{-7}$ torr. Next, the ignitrons of each bank are simultaneously triggered at $t = 0$ to initiate breakdown (the main banks can also be triggered at different times if desired). Because the gas cloud is approximately azimuthally symmetric after leaving the azimuthally symmetric gas slits, gas breakdown forms "Bundt cake", half-donut arched plasma formations (Figs. 10c and 3c leftmost panel). This initial formation is an azimuthally symmetric version of the discrete "spider leg" arches of the Caltech experiment (Hsu & Bellan 2002, 2003; Bellan et al. 2005; You et al. 2005). As current increases, the plasma arches expand to form a collimated high density jet (Figs. 10d, 10e, 3c panels, 7, and 8) similar to the jets from the Caltech experiments (You et al. 2005). The typical total gun current (skin and core guns) is 340 kA for guns charged to -4 kV up to 470 kA when charged to -10kV. Measurements from the interferometer (Fig. 8) show that plasma line-integrated density rises rapidly to $\sim 8 \times 10^{20}$ m$^{-2}$. From high-speed camera images, the plasma jet at that time has ~5 cm diameter so the plasma density is $\sim 1.5 \times 10^{22}$ m$^{-3}$. These observations confirm those of the Caltech jet experiment (Kumar & Bellan 2006; Yun et al. 2007) and reinforces the MHD model for pumping plasma into current-carrying magnetic flux tubes (Bellan 2003; You et al. 2005). The evolution of the plasma jet can be tracked in a stability space (Fig. 8c) defined by an aspect-ratio $\bar{k} \equiv 2\pi a/L$ (where $a$ is the jet radius and $L$ is the jet length) and a normalised current $\bar{\lambda} \equiv \mu_0 I a/\psi$ (where $\mu_0$ is the magnetic permeability, $I$ is the plasma jet current, and $\psi \sim 3.4$ mWb is the bias coil magnetic flux). The plasma jet starts off as a short fat current-carrying magnetic flux tube and lengthens into a long thin current-carrying magnetic flux tube (Figs. 3c, 8a, 10). The $\bar{k}$-$\bar{\lambda}$ stability space is characterized by kink and sausage stability thresholds (von der Linden & You 2017). In static ideal MHD models, the stability thresholds are Kruskal-Shafranov (Kruskal & Tuck 1958) and Tayler sausage instability (Tayler 1957) thresholds. For more general ideal MHD models with finite length and arbitrary current profiles, the stability has thresholds that are modified from the classical Kruskal-Shafranov and Tayler criteria but retain the general behavior: a current-carrying magnetic flux tube starts off stable on the top left of the stability space (Fig. 8c), lengthens and collimates, travelling towards the bottom right of the stability space, goes kink unstable when crossing the Kruskal-Shafranov threshold (Fig. 8a at 12 $\mu s$, Fig. 10f) and undergoes a reconnection due to a Rayleigh-Taylor instability (Moser & Bellan 2012) or a sausage instability (von der Linden & You 2017) (Fig. 8a at 12.5 $\mu s$, Fig. 10g). The plasma jet then loses integrity, turning into an isolated magnetized plasmoid that propagates ballistically (presumably). At this point in time, significant power remains in the power supplies (Fig. 8b after 20 $\mu s$) but uncontrolled arcs between electrodes (Fig. 8b frames 6-12) syphon current away from a new jet formation. The plasma density measured just in front of the electrodes (Fig. 8f) has dropped rapidly due to the broken jet travelling past the laser beam, then rises again slowly as the uncontrolled arcs produce a plasma exhaust plume from teflon ablation mixed with gas plenum ionization (Fig. 8b frames 6-12).

The first operational campaign awaited final commissioning of spectroscopic and magnetic diagnostics, so high speed sequence of images (Fig. 8a) provided a first estimate of the plasma jet velocity of ~80 km/s as it expands into the vacuum chamber, assuming the visible emission is representative of the plasma density. The total gun current in shot #6580 (Fig. 8d) is about 60 kA at 12 $\mu s$. If assumed to flow entirely inside a plasma jet of ~5 cm diameter, the toroidal magnetic field is then ~0.5 T giving an Alfvén velocity of ~86 km/s for the measured plasma density of $1.5 \times 10^{22}$ m$^{-3}$. The jet flow velocity is thus at the Alfvén velocity as expected from the MHD pumping theory (Bellan 2003). Time-of-flight measurements of magnetic field, assumed to be convected with the plasma jet, from magnetic probes commissioned during the second operational campaign resulted in velocities 20-70 km/s. The first spectroscopic Doppler measurements obtained during the third operational campaign (Fig. 9c) on long plasma jets (Fig. 9b) show line-integrated axial flows up to 50 km/s and azimuthal flows >20 km/s. The gun currents and voltage traces provide a rough estimate of the total plasma impedance of ~25 mΩ for the core plasma and ~100 $m\Omega$ for the skin plasma giving temperature estimates of 1-9 eV from Spitzer resistivity. These back-of-the-envelope estimates indicate that the core plasma temperature rises from ~5 eV to ~9 eV while the skin plasma starts colder at ~1 eV to increase by ~10× to match the core temperature as the jet forms, collimates and expands. The plasma inductance has been estimated $l(t) = B^2 \pi a^2 L/(\mu_0 I^2) \sim 30 - 80$ nH for an inductive impedance of $Z_L = 2\pi l/\tau \sim 4 \, m\Omega$ from the observed jet dimensions ($a \sim 3cm, L \sim 20 - 50cm$), magnetic fields ($B \sim 0.5 \, T$) and gun currents ($I \sim 60 \, kA$ with RLC ring period $\tau \sim 140 \, \mu s$) averaged over several discharges. This suggests that the total impedance of these cold jets is presently dominated by plasma resistivity and not by plasma inductance (neglecting electrode sheath effects and other impedances).

The results of this first operational campaign informed an upgrade to the the plasma gun that resulted in significant improvement in plasma jet lifetime (from ~10 $\mu$s to ~100 $\mu$s with the same discharge parameters) during a second operational phase between Dec. 2016 and Mar. 2017, and after the completion of the multichord fiber-bundle and magnetic probe arrays, a third operational phase between Apr.-Jun. 2017. The upgrade replaced the teflon sheet with an alumina ceramic insert to prevent arcing between the middle and outer electrodes. A second plasma jet (Fig.11a, frames 6-12) thus forms after the first one becomes unstable and loses integrity (Fig. 11a, frames 1-5). All the remaining electrical power is then available for driving this second jet. The jet triples in length, reaching the other end of the vacuum chamber at 1.1m away (Fig. 11a, frames 8-10), and remains stable with strong helical flows (Fig. 9). The plasma density measured at the center of the vacuum chamber (Fig. 11c) shows that the second long-lived plasma jet is similary dense to the first short-lived plasma jet ($\bar{n}_e \sim 8 \times 10^{20} \, m^{-2}$) while carrying twice the total current (Fig. 11b, times of frame 7-10 compared to times of frames 1-3). The operational sequence thus continues after the detachment of the first short-lived plasma jet (Fig. 10g). This detached plasma travels into the vacuum chamber (Fig. 11a, frames 3-6, and Fig. 10) while a second jet forms into the plasmoid (Fig. 10h, Fig. 11, frames 5-6) and lengthens to form a long-lived, stable, jet-like flowing current-carrying magnetic flux tube (Figs. 10i, j, k, Fig. 11, frames 7-12).

Why was the second jet not observed earlier? We do not yet have a satisfying answer to this question. The arcing between electrodes is the primary reason, but not the complete answer. Without the insulator between the inner and middle electrodes, a second jet is never obtained. Any arcing is strongly asymmetric and uncontrolled during or after the first short-lived jet. With a Teflon insulator, we do not get the long-lived second jet either. Instead, there is an azimuthally symmetric glow around the insulator associated with Teflon, and a non-collimated plume of hydrogen mixed with Teflon (Fig. 8b, frames 6-12). Spectroscopic observations confirm strong carbon and fluorene presence at this time. In this case, the LabJet gun basically operates as a large pulsed plasma thruster. The plume is not a collimated jet presumably because the glow over the insulator ("skin" plasma) syphons most of the electrical current away from the plume ("core" plasma). Without electrical current, there can be no pumping nor collimation effect. With an alumina insulator, the long-lived second jet forms and there is no glow at the electrodes (arcs occasionally appear), and all the current can now flow into the second jet to drive helical flows.

Preventing arcing between the electrodes is therefore a necessary but, we believe, not a sufficient condition to produce the second jet. The second jet was never observed in the Caltech experiment, despite having similar geometries and capacitor banks, and later, pulse forming networks (Moser & Bellan 2012). The key is to also include another electrode and have azimuthally-symmetric mass sources. Both conditions are necessary to drive helical shear flows sufficiently strong to stabilize such a long jet. Yet, these conditions do not explain why we have a short-lived unstable jet in the first place, and only the second jet becomes long and stable with helical flows. Why did the first jet not become stable and long (as was expected when conceiving the project)? We speculate this is related to the speed at which a jet forms into the vacuum chamber and, possibly, the cross-over of core currents and skin currents (Fig. 4a at 54 $\mu$s and Fig. 11 at 40 $\mu$s). The first jet propagates into vacuum on a few Alfvén times, driven by the Alfvénic axial flows of the gobble theory (Bellan 2003, You et al 2005). So, the jet will reach the Kruskal-Shafranov length on a few Alfvén times, whereupon kink instabilities develop on a few Alfvén times. Helical flows driven by $\boldsymbol{E} \times \boldsymbol{B}$ develops on drift time scales so does not have sufficient time to develop and stabilize the column. Once the first jet becomes unstable and detaches, the second jet propagates not into vacuum but into this finite mass plasmoid. This gives time for helical flows to develop and stabilize the jet before kink instabilites develop. Detailed quantitative results and analysis will be reported in a subsequent paper.

## 8. SUMMARY

The Mochi experiment is thus designed to investigate the interaction between helical magnetic fields and flows in cylindrical geometry by simulating an astrophysical jet launched by an accretion disk with the LabJet configuration. The primary novelty of the setup consists of three electrodes, each with azimuthally symmetric gas slits. Two of the electrodes are biased independently with respect to the third electrode to control the radial electric field profile across the poloidal bias magnetic field, and thus approximates a shear azimuthal rotation profile in an accretion disk. The azimuthally symmetric gas slits in each electrode provides a continuously symmetric mass source at the footpoint of the plasma jet. This design is so azimuthal rotation of the plasma jet is not hindered by a discrete number of gas holes. The combination of a controllable electric field profile, a flared poloidal magnetic field, and azimuthally symmetric mass sources results in a plasma jet made of controlled sheared helical flows in a lengthening, magnetized, current-carrying magnetic flux tube. The measured parameters of the first plasmas are ~$10^{22}$ m$^{-3}$, ~0.4 T, 5-10 eV, with velocities of 20-80 km/s successfully producing short-lived (~10 $\mu s$, $\gtrsim$ 5 Alfvén times) collimated jets with ~10:1 aspect-ratio and long-lived (~100 $\mu s$, $\gtrsim$ 40 Alfvén times) flow-stabilized, collimated jets with ~30:1 aspect-ratio. Vector tomographic reconstruction of flow fields from spectroscopic measurements are in the preliminary stages. The theoretical framework for investigating the evolution of flows and magnetic fields is based on the self-organization of canonical flux tubes. The Mochi experiment can also be adapted to operate both guns simultaneously for studies of jet merging or jet impact on targets to simulate astrophysical jet impacts on vacuum magnetic fields, on neutral gas or on plasma clouds.

The work was supported by the U.S. DOE Early Career Award DE-SC0010340. J. von der Linden was partially supported by the U. S. DOE Office of Science Graduate Student Research Program. The authors acknowledge K. Vereen, Y. Kamikawa, J. Stuber, R. Golingo, U. Shumlak, T. Mattick, P. Bellan, S. Hsu, S. Woodruff, G. Yun, Y. Ono, and the undergraduates T. Coleman, R. Pratt, P. Rudolf, J. Wiegman, D. Crewes, C. Cretel, J. Geier, D. Cho, D. Lim, for contributions to the project, and the University of Washington Student Technology Fund for loan of the Kirana camera.

**Table 1**
Parameters of the Mochi.LabJet experiment

| (a) Mochi Machine Hardware Name | Design | Operation to-date |
|---|---|---|
| LabJet.Gun.InnerElectrode | 7.25 cm outer diam. Cu disk, 6.1 cm diam. at gas slit | PSU2 connects inner to outer electrode |
| LabJet.Gun.MiddleElectrode | 17.8 cm outer diam. Cu annulus, 9.8 cm diam. at gas slit | PSU3 connects middle to outer electrode |
| LabJet.Gun.OuterElectrode | 90cm outer diam. Cu annulus, 38.2 cm & 80.2 cm diam. at gas slits | Common electrode |
| SpiderLeg.Gun.InnerElectrode | 49.6 cm diam. W-coated Cu disk, 48.26 cm diam. at gas holes | PSU1 connects inner to outer electrode |
| SpiderLeg.Gun.OuterElectrode | 90 cm outer diam. W-Cu annulus, 83 cm diam. at gas holes | Common electrode |
| LabJet.BiasFieldCoil using PSUa | 99 turns, 13 cm diam., 1 mH, 1–5 mWb | 0.1 – 0.37 T, 1.5 – 5 mWb |
| LabJet.BiasFieldCoil using PSUb | 99 turns, 13 cm diam., 1 mH, 1–5 mWb | 0.2 – 0.52 T, 2.8 – 6.9 mWb |
| Vacuum chamber | Spherical, 1.4 m diameter, electropolished stainless steel, | $3\times10^{-7}$ torr with cryopump |
| Ports | 2×41" wire-seal, 2×14" Conflat, | Plasma guns, pumps, blank |
| Viewports | 6×8" equatorial, 64×3.375" spherical radial, 30×2.75" planar radial | Cameras |
| Gas | H, D, He, Ar, N, Ne, Kr | H, Ar, N |
| PSU1: for SpiderLeg.Gun | 12 kJ, 240 $\mu$F, 10 kV | 4 kV |
| PSU2: for LabJet.CoreGun | 30 kJ, 600 $\mu$F, 10 kV | 4 – 10 kV, 95 – 235 kA |
| PSU3: for LabJet.SkinGun | 30 kJ, 600 $\mu$F, 10 kV | 4 – 10 kV, 90 – 233 kA |
| PSUa: for bias magnetic field | 12 kJ, 600 mF, 200 V | 50 – 200 V, 180 – 600 A |
| PSUb: for bias magnetic field | 6 kJ, 48 mF, 500 V | 195 – 490 V, 350 – 850 A |
| **(b) Mochi.LabJet Plasma Parameters** | **Design** | **Measured to-date** |
| Plasma lifetime | 10 – 50 $\mu$s | 4 (first jets) – 70 (final jets) $\mu$s |
| Density | $10^{17} – 10^{23}$ m$^{-3}$ | $10^{22} – 10^{23}$ m$^{-3}$ |
| Temperature | 1 – 10 eV | 5 – 10 eV |
| Magnetic field | 0.01 – 0.4 T | 0.4 T |
| Axial velocity | 50 – 100 km/s | 40 – 80 km/s |
| Alfvén velocity | 10 – 1,000 km/s | 20 – 80 km/s |
| Radius | 4 – 10 cm | 2 – 4 cm |
| Length | 50 – 100 cm | 20 – 110 cm |
| Pressure | $0.02 – 2\times10^6$ Pa | $10^3 – 10^5$ Pa |
| Ion skin depth | $\leq 1$ cm | 0.02 – 0.07 cm |
| Lundquist number | 10 – 10,000 | 30 – 300 |
| Plasma $\beta$ | 0.001 – 0.5 | 0.1 - 0.3 |
| Mach number | 1 – 15 | 1 – 4 |
| **(c) Comparison to protostellar jet** | **Mochi.LabJet;** | **Protostellar jet;** | **Ratio** | **Symbol for ratio or scaling expression** |
| Diameter | 0.1 m; | $10^{15}$ m; | $10^{16}$ | $a$ |
| Mass density | $10^{-4}$ kg/m$^3$; | $10^{-17}$ kg/m$^3$; | $10^{-13}$ | $b$ |
| Pressure | $3\times10^4$ Pa; | $10^{-9}$ Pa; | $4\times10^{-14}$ | $c$ |
| Time scales | 50 $\mu$s; | 10000 yrs; | $6\times10^{15}$ | $d$ |
| Velocity scales | 100 km/s; | 300 km/s; | 3 | $e$ |
| Magnetic field scales | 0.5 T; | $10^{-7}$ T; | $2\times10^{-7}$ | $f$ |
| MHD scaling: time | 2.5 | | | $a\sqrt{b/c}\,/\,d$ |
| MHD scaling: velocity | 0.2 | | | $\sqrt{b}\,c\,/\,e$ |
| MHD scaling: magnetic field | 1.0 | | | $\sqrt{c}\,/\,f$ |

**Notes.** The label LabJet stands for the new triple-electrode plasma gun described in this paper; SpiderLeg stands for the gun on the opposite end of the chamber, unused and floating in this paper; PSU stands for Power Supply Unit (our capacitor banks); CoreGun stands for the gun made of the inner and outer electrode pair driving the core of the jet; SkinGun stands for the gun made of the middle and the outer electrode pair driving the skin of the jet; BiasFieldCoil for the external coil imposing an initial vacuum dipole magnetic field.

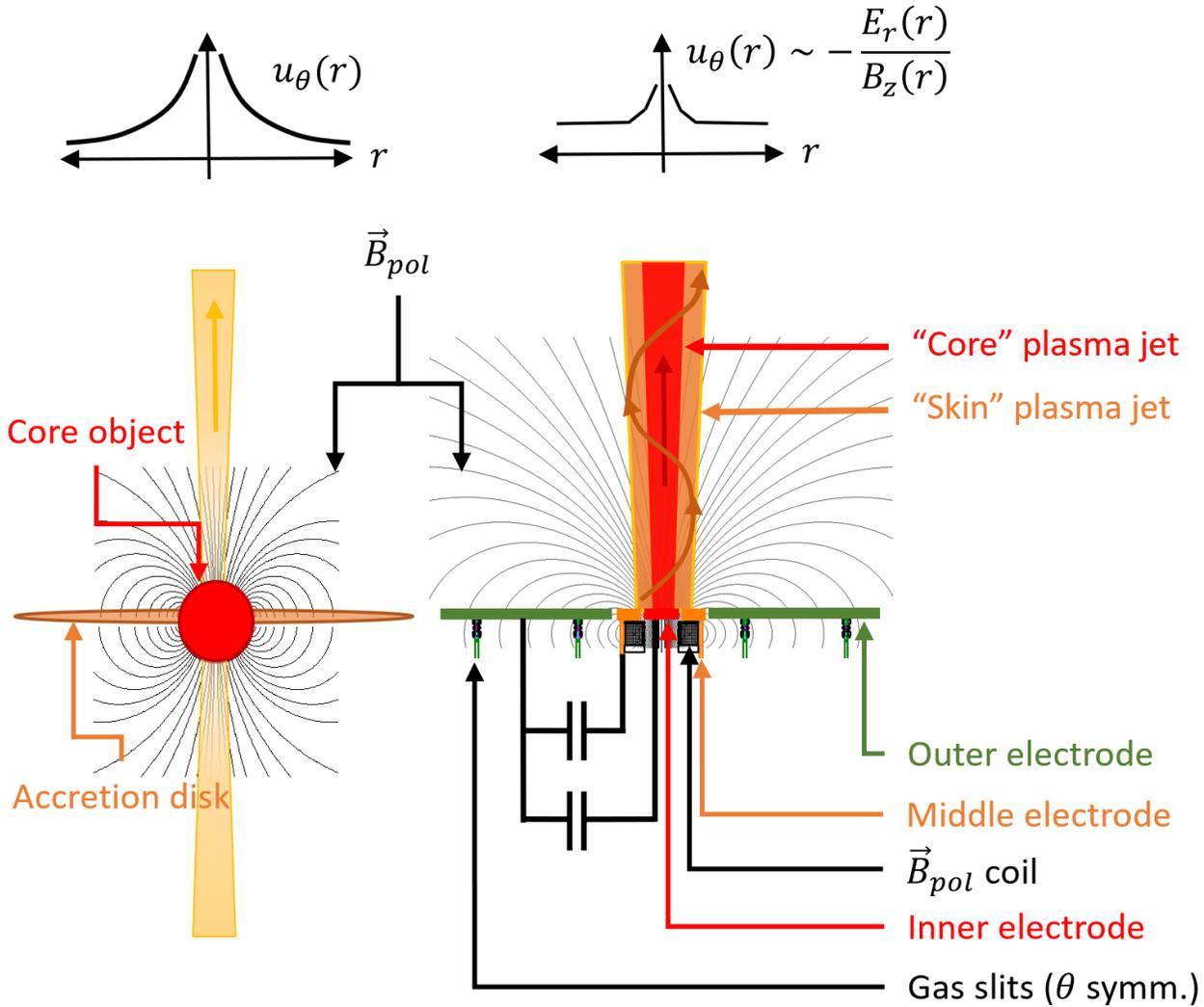

**Figure 1.** Concept behind the LabJet configuration of the Mochi facility: the setup replicates the basic configuration of an accretion disk rotating around a central celestial object with a magnetic field (left) with three planar disk and annular electrodes threaded by a poloidal magnetic field (right). The three electrodes represent the minimum number of electrodes necessary to have a non-uniform radial electric field $E_r(r)$ that, combined with a poloidal magnetic field $B_z(r)$, mimics a non-uniform azimuthal rotation profile $u_\theta(r) \sim -E_r(r)/B_z(r)$ in an accretion disk. The inner electrode is designed to drive the core of the plasma jet and the middle electrode to drive the outer skin of the jet, and therefore to provide some degree of control on the jet current and flow profile. The mass source into the jet footpoints are azimuthally symmetric gas slits at the electrodes, as opposed to discrete numbers of gas holes, to allow for free azimuthal rotation.

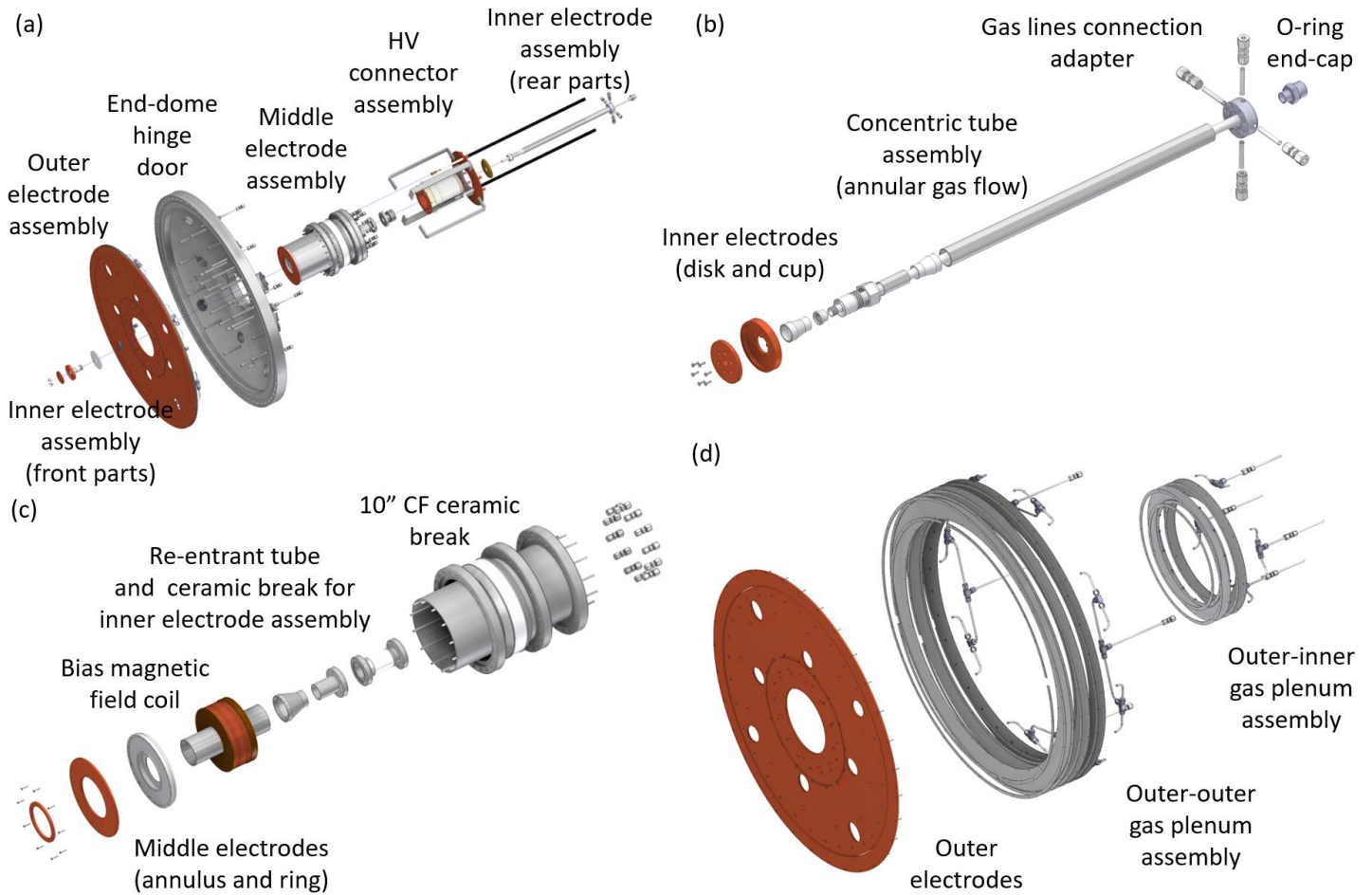

**Figure 2.** (a) The Mochi.LabJet plasma gun assembly showing how the sub-assemblies are mounted together on the end-dome door of the vacuum chamber; (b) the inner electrode sub-assembly with concentric tubing, to provide azimuthally-symmetric gas out the slit between the disk and cup copper ends, and provide a clear axial viewpoint for future diagnostics inserted through the end cap connector; (c) the middle electrode sub-assembly that houses the bias magnetic field coil and the inner electrode sub-assembly with a ceramic break and an azimuthally-symmetric gas plenum behind the copper electrodes; (d) the outer electrode sub-assembly consisting of copper electrodes connected to azimuthally-symmetric gas plenums made from stacks of slotted stainless-steel sheets; the eight cutouts on the electrodes are for optical view access from the chamber viewports.

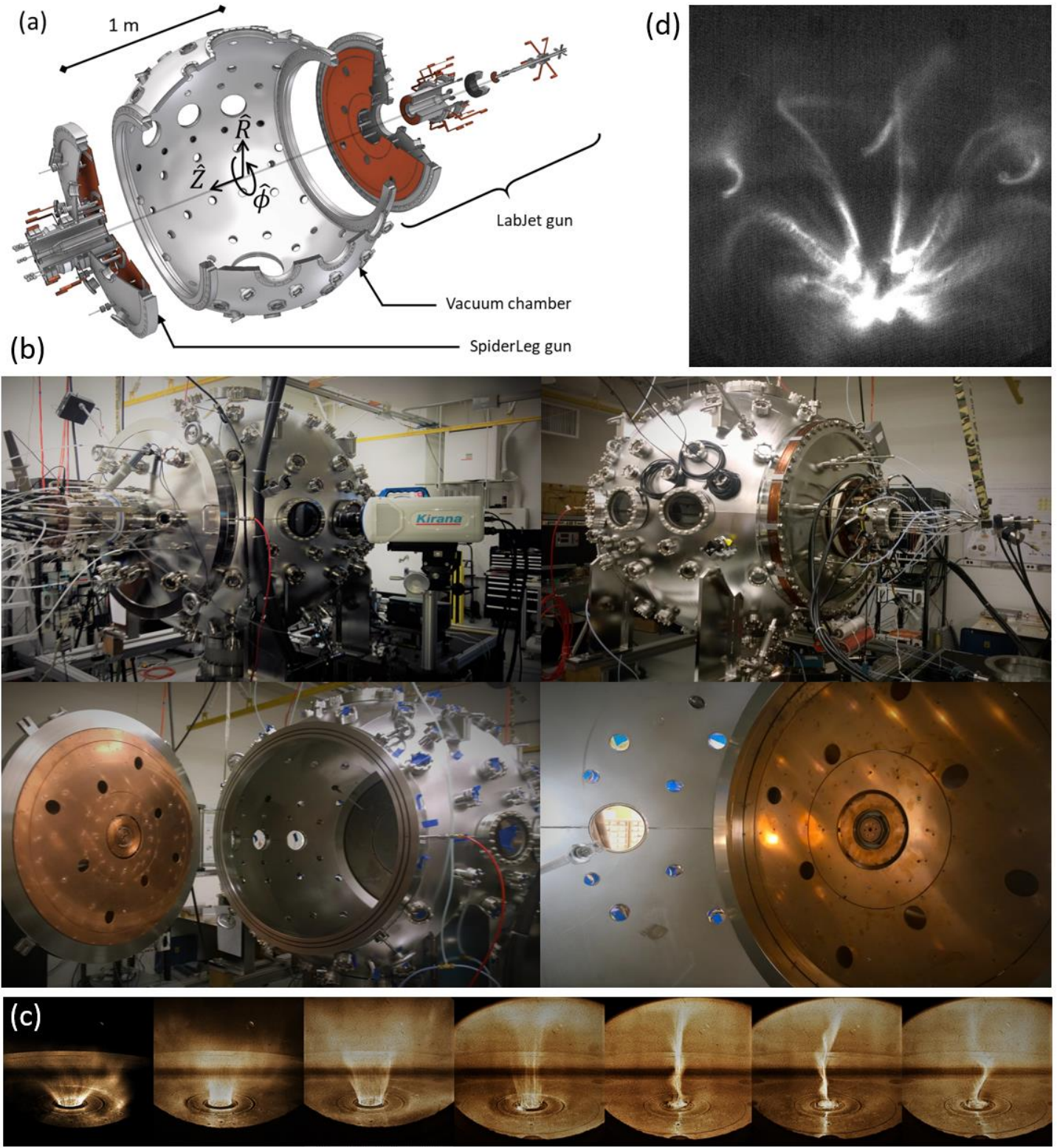

**Figure 3**. The Mochi facility: (a) schematic showing the mounting of the LabJet gun opposite the SpiderLeg gun on the spherical vacuum chamber. The numerous viewports generally point radially to the center of the sphere, except for several diagnostic ports that point to the Z-axis of the chamber within an azimuthal plane. The cutouts in the LabJet gun provide access to the viewports behind the gun, there are none behind the SpiderLeg gun; (b) photographs of the apparatus showing the LabJet gun (top left), the SpiderLeg gun (top right), the hinged door with the LabJet gun and the tungsten coated SpiderLeg gun in the background (bottom left), the typical high-speed camera view of the LabJet gun (bottom right); (c) Typical plasma jets formed from the LabJet gun during the first operational campaign showing azimuthally-symmetric tubes forming, lengthening, collimating and undergoing instabilities; (d) Early plasma discharge in the SpiderLeg gun showing discrete, kinked, current-carrying magnetic flux tubes linking the gas holes.

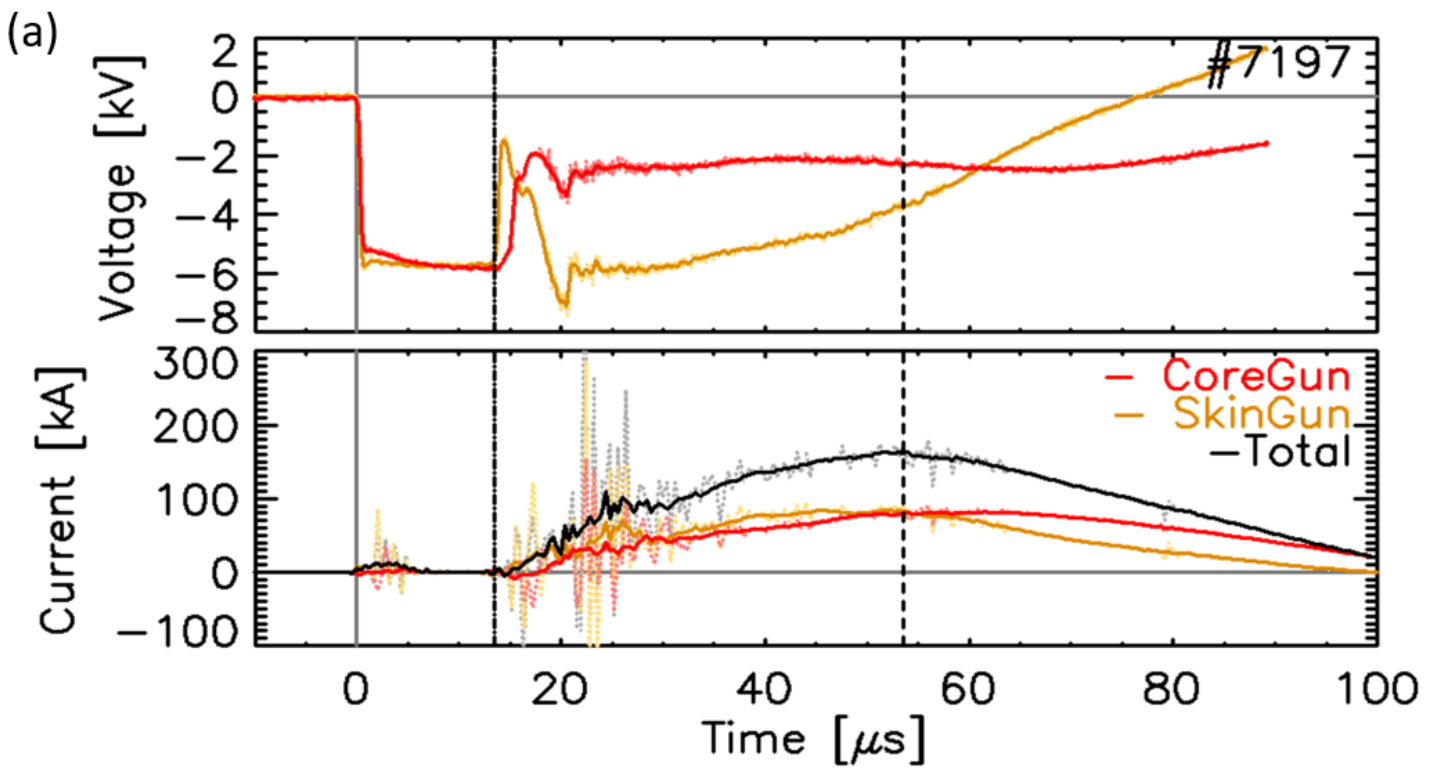

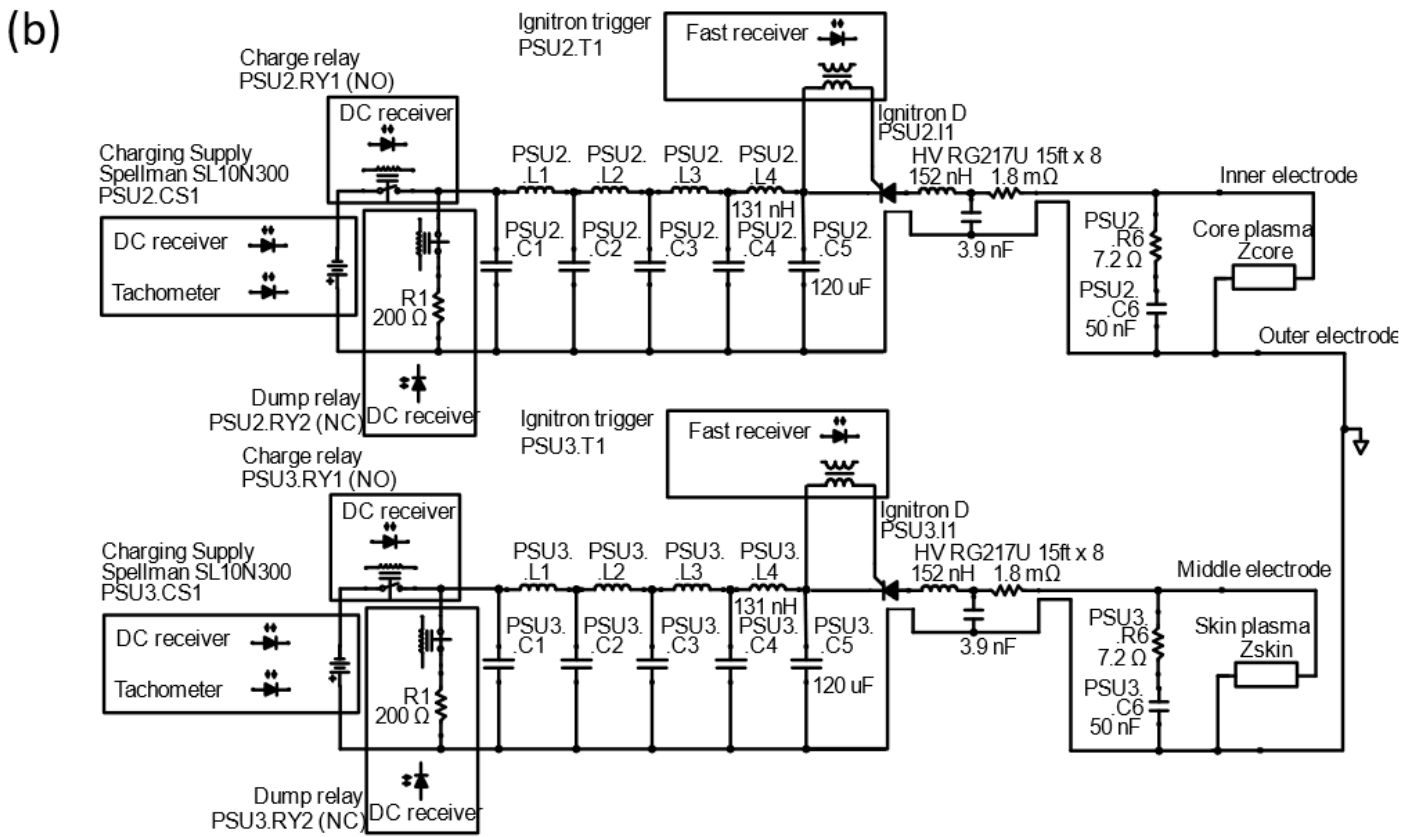

**Figure 4.** Custom power supplies and typical discharge of the Mochi.LabJet experiment: (a) voltage and current traces during a typical discharge, showing separate evolutions of the currents and voltages measured at the electrodes; the dash-dot line represents the gas breakdown time, the dashed line represents the "crossover" time when the CoreGun current becomes larger than the SkinGun current; (b) one of the custom power supplies assembled into an enclosure; (c) circuit diagram of the experiment with remote operation of the relays, charging demand, and triggering with custom opto-electronic isolation units (Figs. 5-6); the assembly can be readily modified into a pulse-forming network with custom inductances between the capacitors; (d) schematic of the custom coaxial ignitron tower assembly.

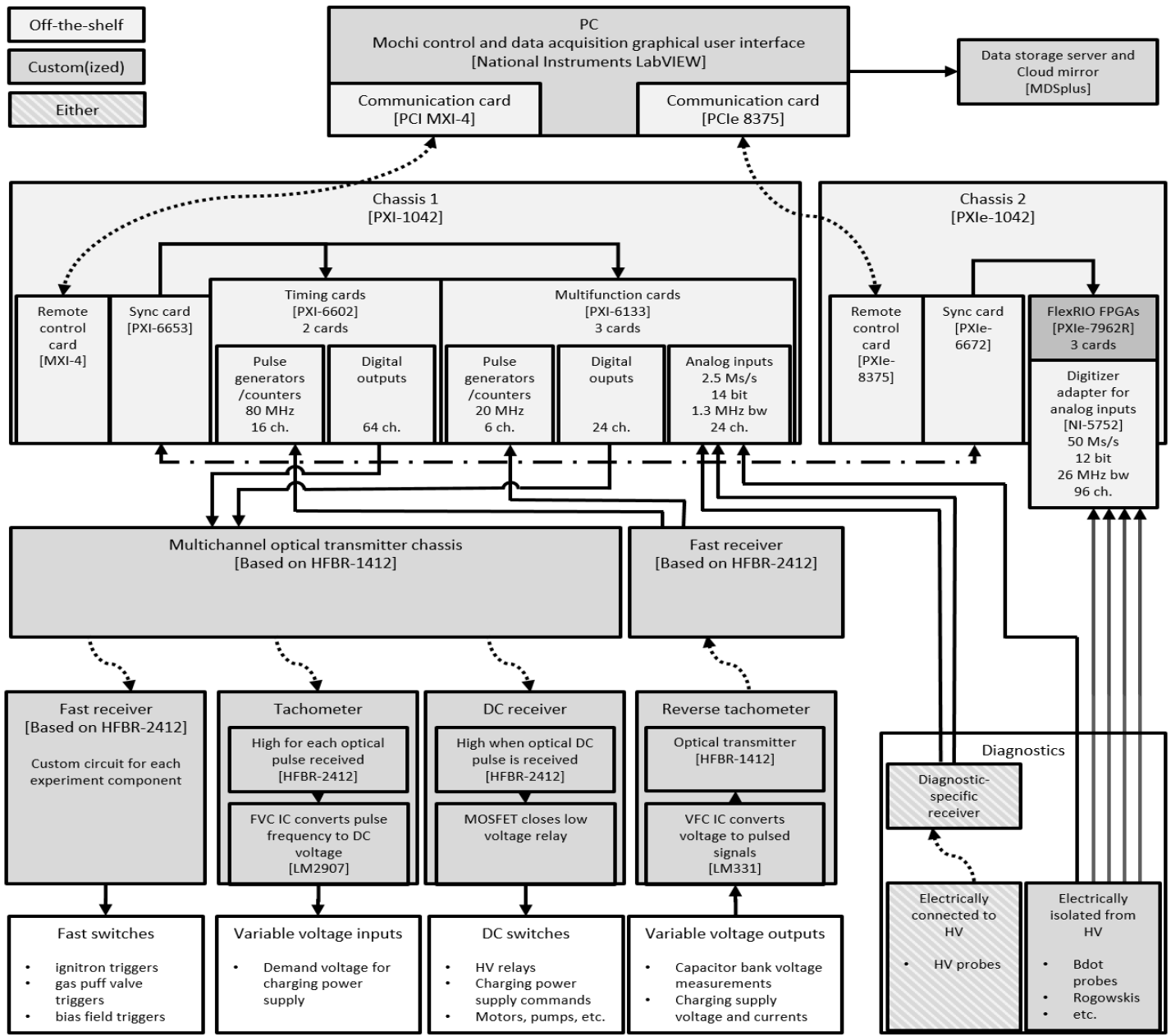

**Figure 5.** Optically-isolated control, feedback, and data acquisition of the Mochi experiment. Solid arrows represent internal or external electrical connections. Dotted curved lines represent fiber-optic connections. The dash-dot line represents a specific connection for synchronization of the two separate chassis. The custom transmitter, receiver, and tachometer circuits are shown in Fig. 6 and control the hardware shown in Fig. 4c.

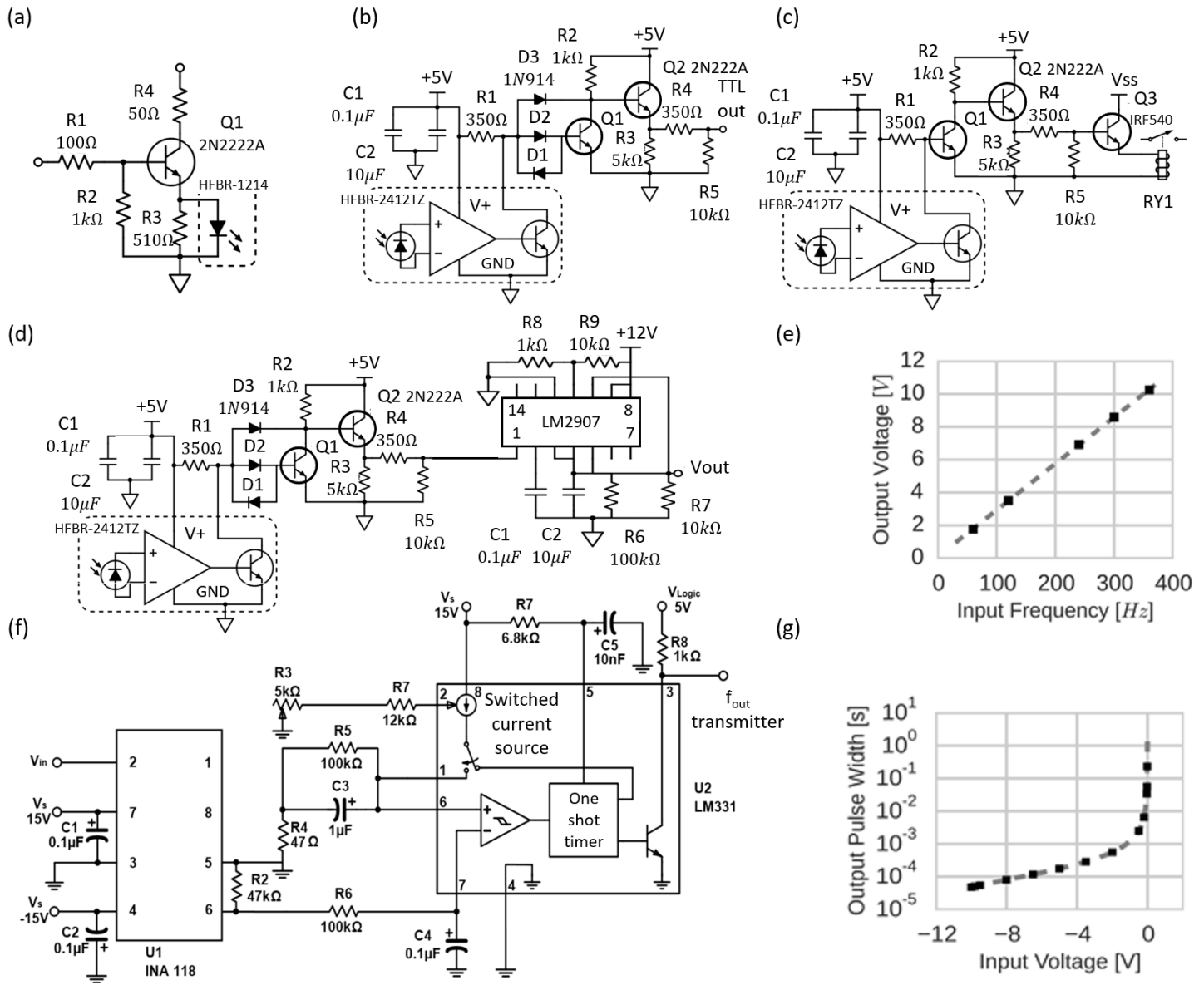

**Figure 6.** Custom opto-electronic circuits: (a) fast transmitter with 310 ns delay, 2.8 ns jitter at 0.1W power consumption; (b) fast receiver with 250ns switching time; (c) DC receiver for actuators and relays; (d) optical tachometer for converting optical frequency pulses to DC output voltage; (e) calibration curve of optical tachometer; (f) reverse optical tachometer for converting input DC voltages to optical pulse widths; (g) calibration curve for reverse optical tachometer.

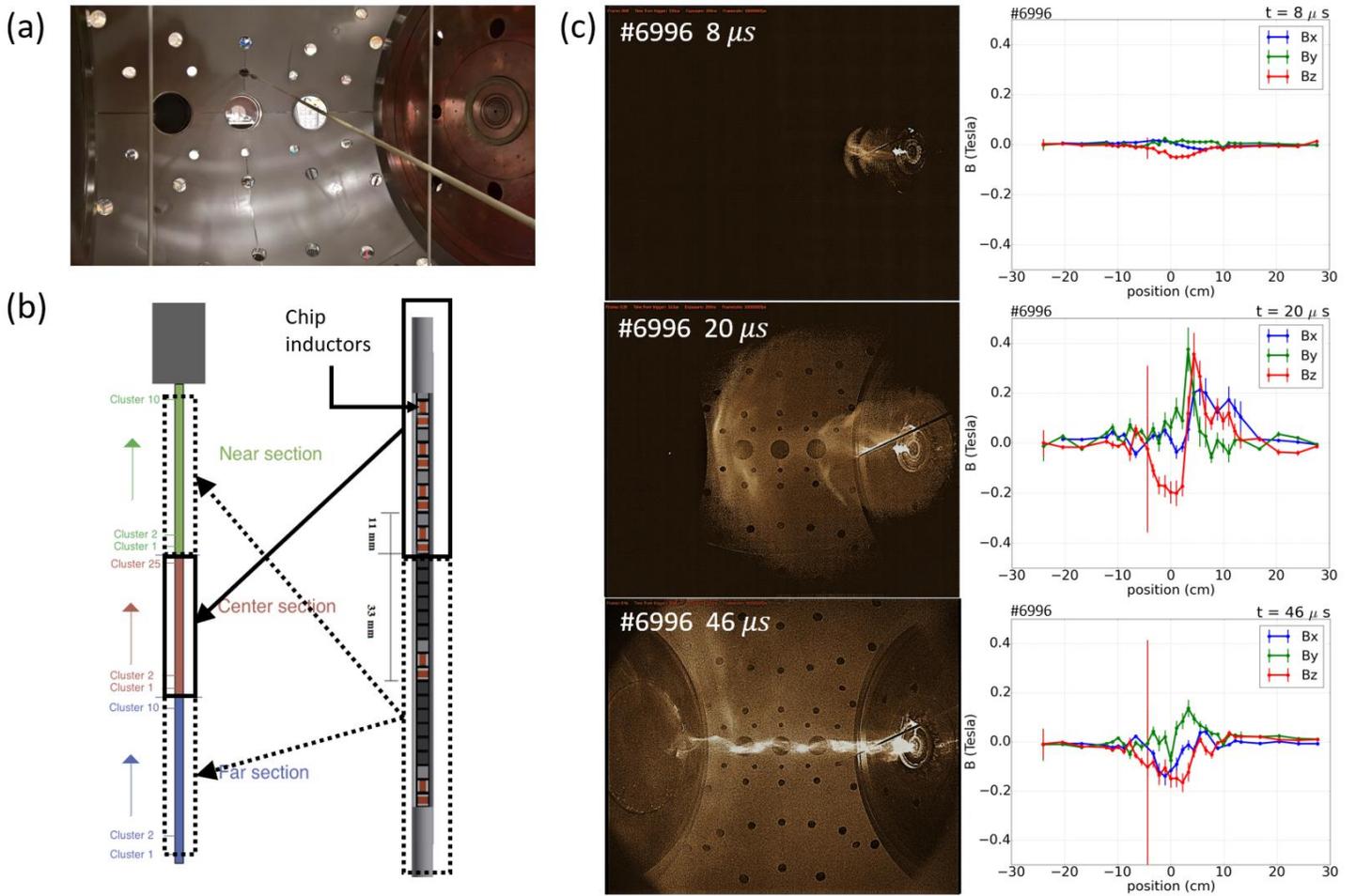

**Figure 7.** Magnetic probe arrays in the Mochi experiment. (a) Three arrays are inserted into the vacuum chamber at select axial locations or any of the other ports available; (b) Each array consists of 135 chip inductor coils arranged into near, center, and far, {x;y;z} clusters; (c) typical magnetic strength vs cluster position data from a single array (0 cm position corresponds to the center of the array) during a plasma discharge in the second operational campaign, showing magnetic fields up to 0.4 T, with profiles resulting from a current-carrying magnetic flux tube.

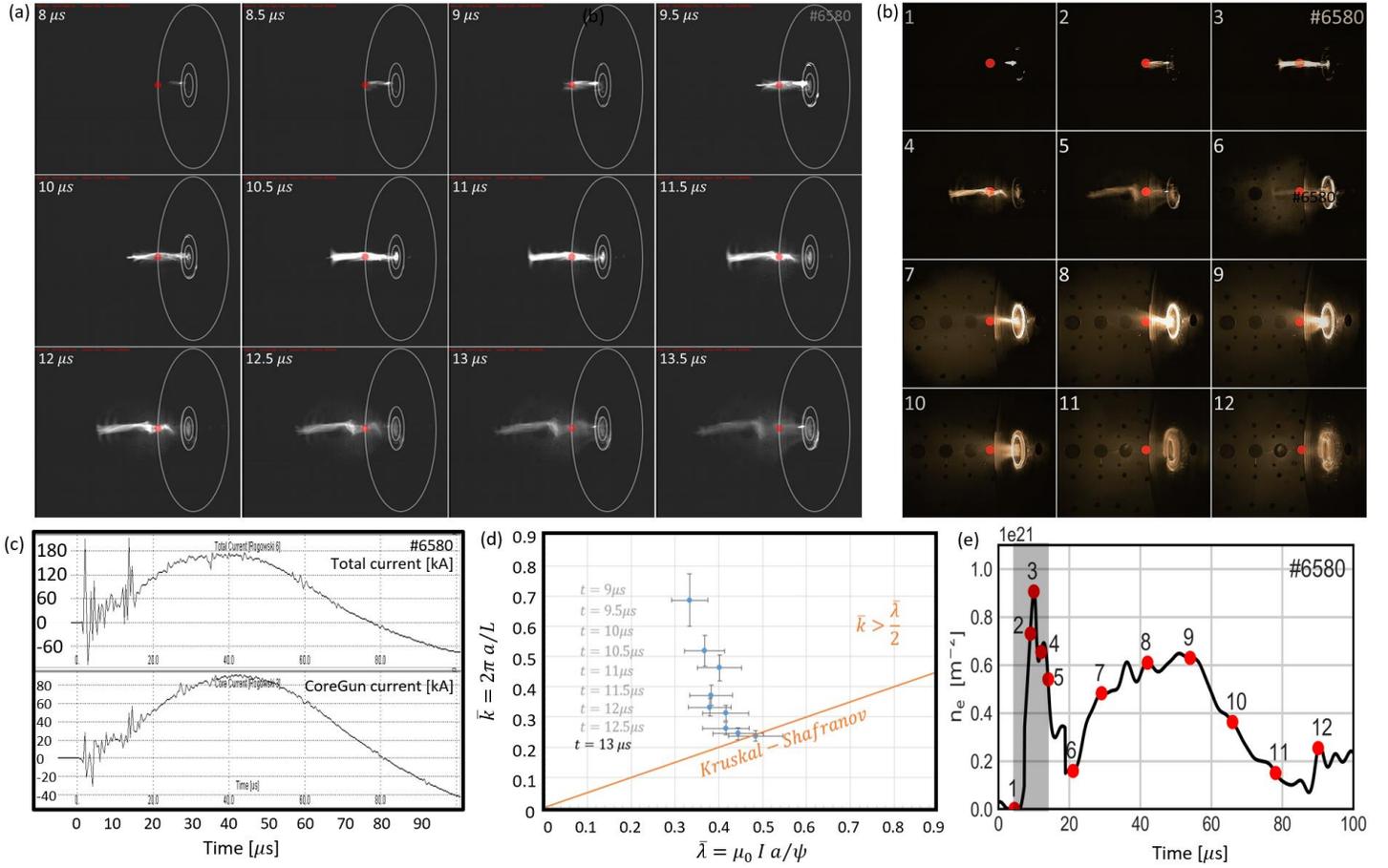

**Figure 8.** Typical evolution of plasma jet during the first operational campaign (shot #6580): (a) formation of collimated plasma jet; ellipses correspond to gun electrodes; (b) photographic sequence of the same discharge over 90 $\mu$s showing a diffuse plume and teflon glow after the jet of panel a disconnects (frame numbers correspond to times in panel e); (c) measured gun currents; (d) evolution in $\bar{k} - \bar{\lambda}$ stability space of the jet in panel a; (e) line-integrated electron density evolution during the plasma jet collimation and expansion, measured at the red spots of panels a and b; the grey time period corresponds to time period of panel a.

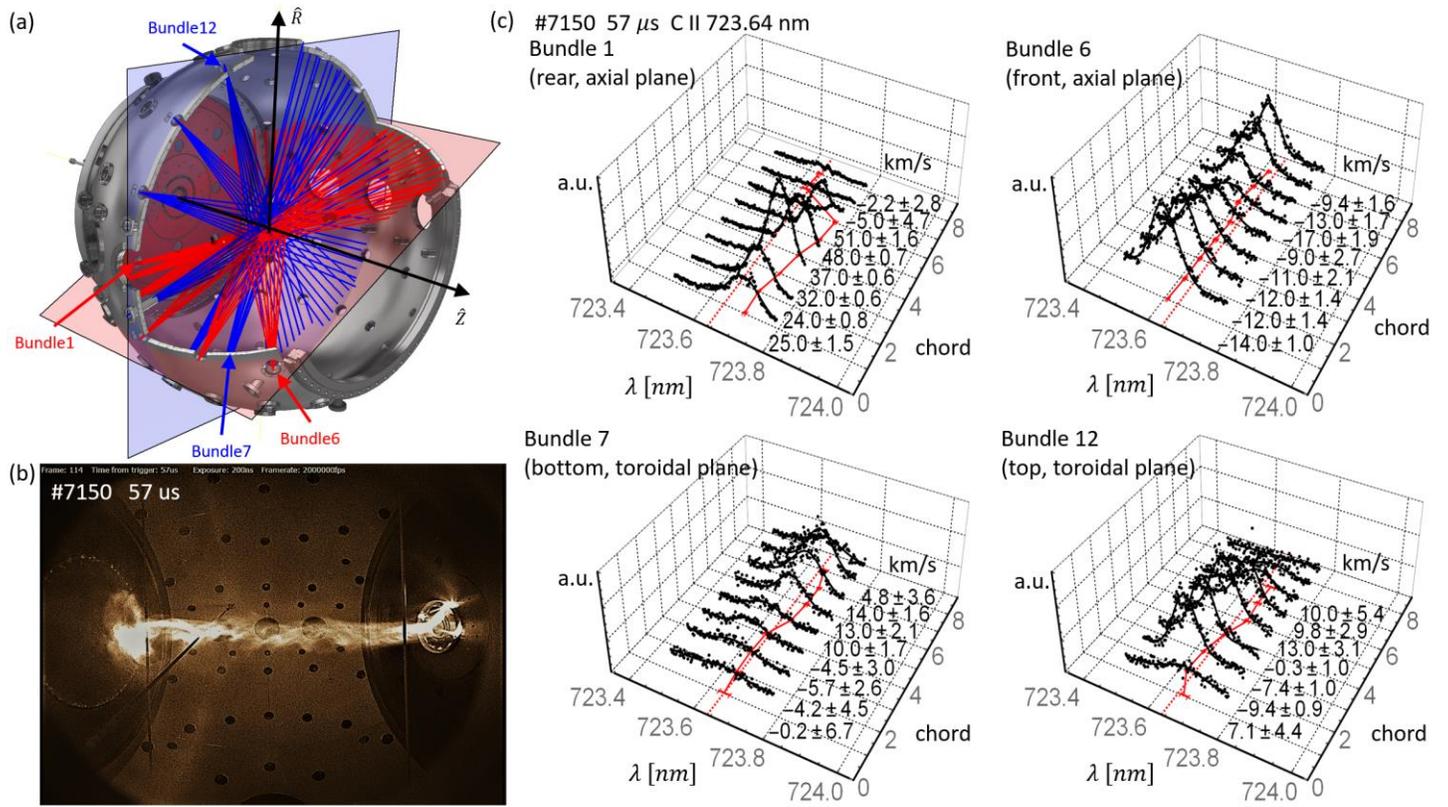

**Figure 9.** Multichannel optical spectroscopy on the Mochi experiment: (a) 96 viewchords arranged in two planes, bundles number 1 to 6 with 8 view chords each in the poloidal axial plane (red), and bundles number 7 to 12 with 8 view chords each in the toroidal plane (blue); (b) fast imaging frame of a stable collimated plasma jet from discharge #7150 at 57 $\mu$s, for which (c) the C II 723.642nm line measurements show a distinct red shift for bundle 1 and a blue shift for bundle 6 indicating axial plasma flows, and bundles 7 and 12 measure blue and red shifts consistent with azimuthal rotation.

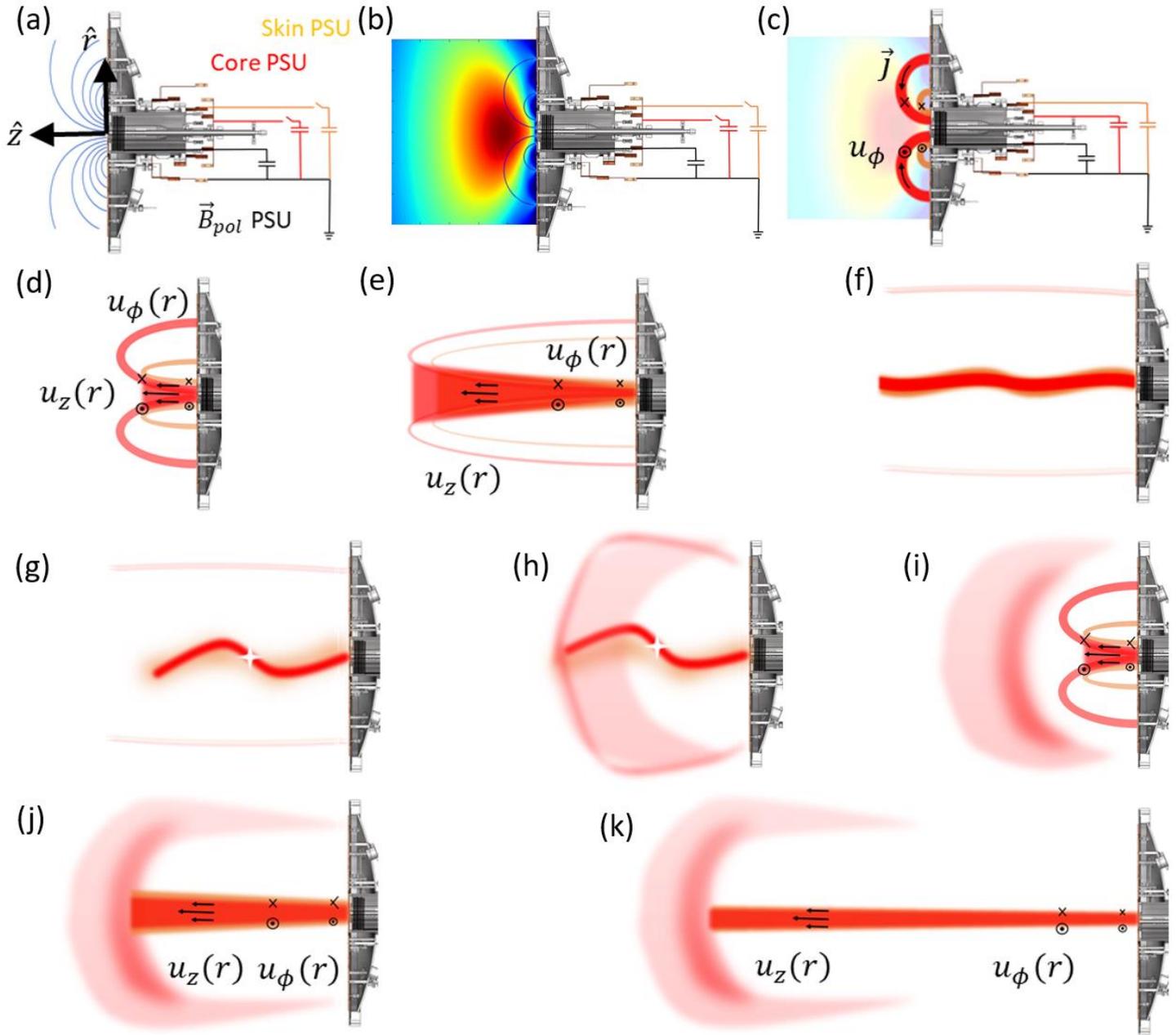

**Figure 10.** Typical operational sequence: (a) a vacuum poloidal magnetic field is applied; (b) gas is injected with fast puff valves to form an azimuthally symmetric gas cloud close to the electrodes; (c) power supply units (PSUs) are switched to drive current through the LabJet.SkinGun (orange) and LabJet.CoreGun (red) pairs of electrodes forming a Bundt cake arched plasma (inset high speed photograph). An azimuthal flow due to a radial electric field crossed with an axial magnetic field develops; (d) a flared current-carrying magnetic flux tube forms, ingesting plasma at high speeds from the footpoints into the vacuum chamber; (e) the plasma jet lengthens and collimates; (f) the plasma jet develops kink instabilities once the Kruskal-Shafranov condition is satisfied; (g) the kinked jet undergoes a reconnection and detaches from the electrodes. During second and third operational campaign, (h) the first jet detaches and propagates into the vacuum chamber (h), (i) a second jet forms and (j) lengthens into the plasmoid (i,j,k) to form (k) a collimated, stable, 30:1 aspect-ratio jet with helical flows.

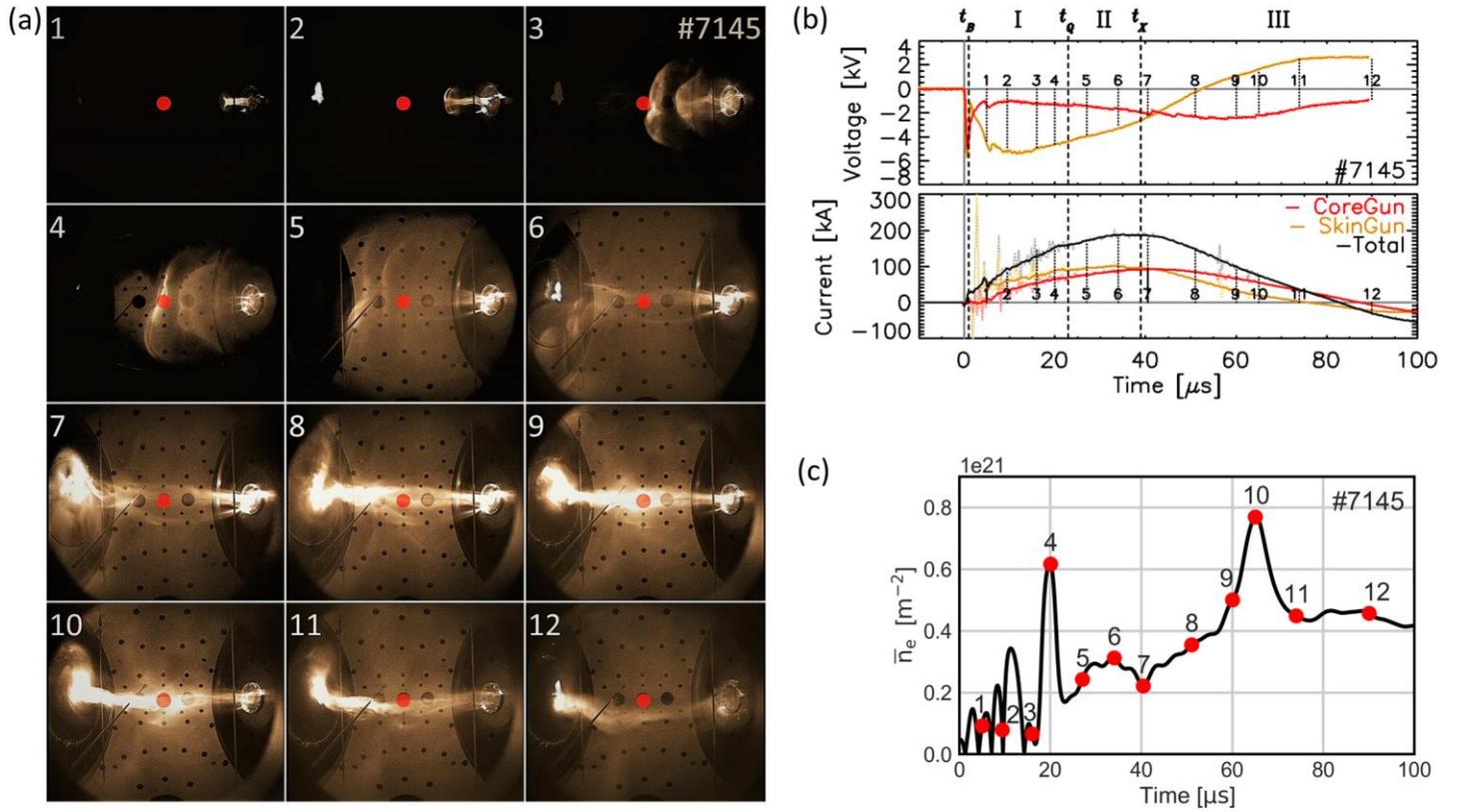

**Figure 11.** Typical operation of second and third operational campaign: (a) visible images from fast-framing camera showing the first short-lived plasma jet (frames 1-3) similar to those of the first operational campaign (Fig. 8) followed by a second formation sequence into the travelling plasmoid (frames 4-6) and collimating into a stable long-lived jet (frames 7-12)); (b) the gun currents and voltages with the times of the frames shown in panel a; (c) the plasma line-integrated density measured in the center of the vacuum chamber (red dot of panel a).